\documentclass[aps,prb,twocolumn,groupedaddress,showpacs,superscriptaddress]{revtex4-1}
\usepackage{graphicx}
\usepackage{color}
\begin{document}
\title{The superfluid density in continuous and discrete spaces:\\ Avoiding misconceptions}

\author{V.G.~Rousseau}
\affiliation{Department of Physics and Astronomy, Louisiana State University, Baton Rouge, Louisiana 70803, USA}

\relpenalty=10000      
\binoppenalty=10000

\begin{abstract}
We review the concept of superfluidity and, based on real and thought experiments, we use the formalism of second quantization
to derive expressions that allow the calculation of the superfluid density for general Hamiltonians with path-integral methods.
It is well known that the superfluid density can be related to the response of the free energy to a boundary phase-twist, or to the
fluctuations of the winding number. However, we show that this is true only for a particular class of Hamiltonians. In order
to treat other classes, we derive general expressions of the superfluid density that are valid for various Hamiltonians.
While the winding number is undefined when the number of particles is not conserved, our general expressions allow us to
calculate the superfluid density in all cases. We also provide expressions of the superfluid densities associated to the
individual components of multi-species Hamiltonians, which remain valid when inter-species conversions occur. The cases of continuous and discrete spaces are discussed, and we
emphasize common mistakes that occur when considering lattices with non-orthonormal primitive vectors.
\end{abstract}

\maketitle

\section{Introduction}
Superfluidity is a manifestation of quantum mechanics at the macroscopic level, and its discovery is usually attributed to
Kapitza\cite{Kapitza}, and Allen and Misener\cite{Allen}.
While superfluidity is widely discussed in the litterature\cite{Fisher,PollockCeperley,Legget,Scalapino,Batrouni,Sorella,Balazs1,Balazs2}, many references quantify this phenomenon by postulating formulae,
or by making semi-empirical derivations. This induces some misconceptions about superfluidity that can lead to mistakes.

The aim of the present paper is first to give a definition of the superfluid density that is based on known experiments.
Then, some expressions of the normal and the superfluid densities that can be used with path-integral methods are rigorously derived from a thought experiment
by using the formalism of second quantization.
We show that, for a particular class of Hamiltonians, the superfluid density
is directly related to the response of the free energy to a boundary phase-twist\cite{Fisher}. While many references improperly use this relation as a general definition of
the superfluid density, we clearly state the condition that the Hamiltonian must meet for such a definition to be meaningful.
We also derive how the free energy is related to the winding number, and recover the expression that was obtained earlier in the context of first quantization\cite{PollockCeperley}.
A drawback associated to the winding number is that it is undefined for Hamiltonians that do not conserve
the number of particles. This problem is common when considering systems with several species of particles where conversions between the
different species occur, and results in the impossibility to calculate the superfluid densities of the individual species.
However, our general expressions of the superfluid density do not rely on the concept of the winding number, and can be used
to determine the superfluid densities of all the species, whether their populations are conserved or not.

The case of lattice Hamiltonians is considered by making a careful discretization of space. We point to some common mistakes
that occur when considering lattices with non-orthonormal primitive vectors. In particular, we show that using the expression of the Laplacian in the natural
coordinates of the lattice requires a change of the energy scale that must be reflected in the expression of the superfluid density.
Also, the metric tensor associated to the natural basis of the lattice must be taken into account when calculating quantities that
involve dot-products, such as the fluctuations of the winding number. As an illustration, in addition to the usual expression of the superfluid density for the \mbox{$d$-dimensional}
cubic lattice, we provide the correct expressions for the triangular, face-centered cubic, honeycomb, kagome, and pyrochlore lattices.
Finally, we give two examples of Hamiltonians for which the well-known expressions of the superfluid density\cite{Fisher,PollockCeperley} are not applicable.
We determine the correct superfluid densities by using our general expressions, and we show that our results are consistent.

\section{Experimental evidences of superfluidity}
\begin{figure}[h]
  \centerline{\includegraphics[width=0.45\textwidth]{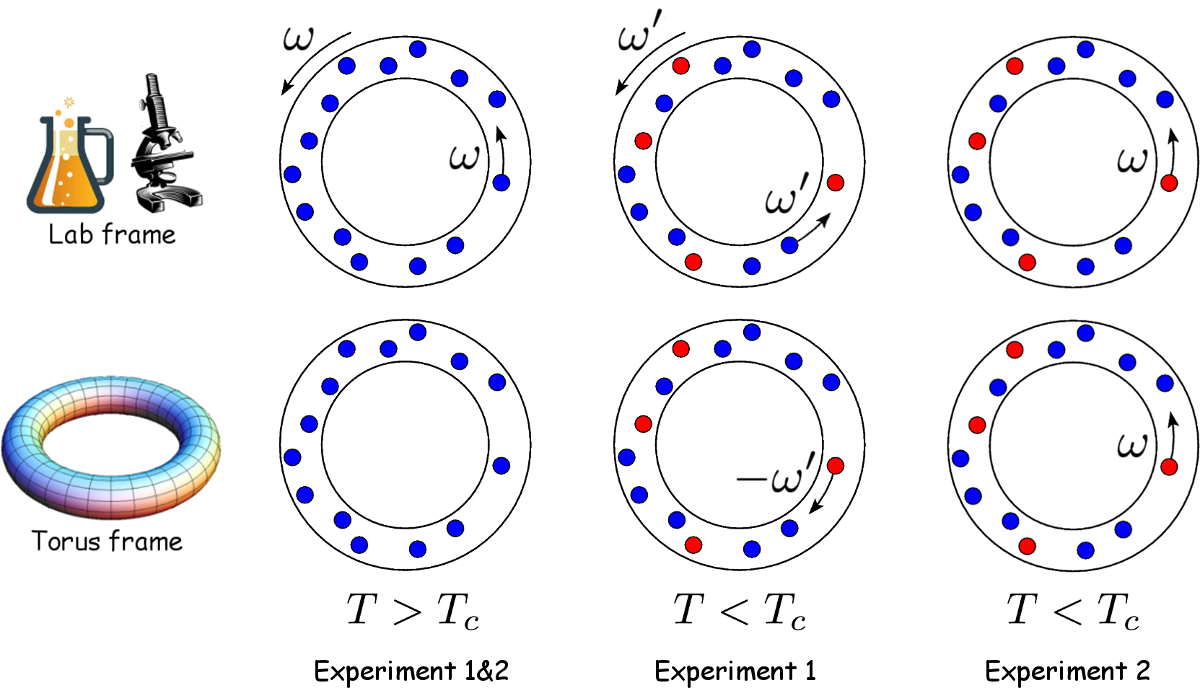}}
  \caption
    {
      (Color online) Analogs of the Meissner effect (experiment~1) and persistent dissipationless currents (experiment~2) observed
      in superconductors. At temperature $T>T_c$, all fluid is in a normal state (blue spheres) and spins at the
      angular frequency of the torus. For $T<T_c$, a fraction of the fluid becomes superfluid (red spheres) and decouples
      from the normal fluid.
    }
  \label{Experiment}
\end{figure}

As suggested by Legget\cite{Legget}, it is useful to consider two experiments that demonstrate fundamental defining properties of a
superfluid (Fig.~\ref{Experiment}):
\begin{itemize}
\item A torus containing liquid Helium $^4$He at temperature $T>T_c$ is spun around its axis at \textit{low} angular frequency,
and left in freewheel. Eventually, because of friction, the liquid comes to equilibrium with the moving walls, resulting in a
constant angular frequency $\omega$ of the torus. Reducing $T$ below $T_c$, an increased angular frequency $\omega^\prime>\omega$
of the torus is observed. The conservation of angular momentum implies that a fraction of the liquid, the superfluid,
decouples from the rest of the liquid, the normal fluid, and spins at lower (possibly zero) angular frequency. This experiment\cite{Hess}
demonstrates the analog of the Meissner effect observed in superconductors.
\item Starting with the same setup as above, the torus is spun at \textit{high} angular frequency. The temperature is then reduced
below $T_c$ and the torus is brought to rest. Eventually the normal fluid comes to equilibrium with the walls. It
can then be verified that the angular momentum of the stationary torus is non-zero (for example, by putting the torus back into freewheel and raising
the temperature above $T_c$, the torus spontaneously starts to spin). The conservation of angular momentum implies that, while the torus
is at rest, the superfluid is still flowing and may continue to do so for a very long time. This experiment\cite{Whitmore,Ekholm}
demonstrates the analog of persistent dissipationless currents observed in superconductors.
\end{itemize}

The outcome of these two experiments can be understood by considering, from the viewpoint of the lab frame, the circulation
of the momentum operator $\vec{\mathcal P}=-i\hbar\vec\nabla$ along a closed loop $\Gamma$ inside the torus around the main axis:
\begin{equation}
  \label{Circulation}\hat\mathcal C=\oint_\Gamma\vec{\mathcal P}\cdot\textrm{d}\vec r
\end{equation}
In the experiment where the torus and the normal fluid are at rest while the superfluid is flowing, the
circulation is due to the superfluid only.
Suppose that the system is in a state $\big|\Psi\big\rangle$ that extends over the volume $\Omega$ of the torus. Then the wave function at point $\vec r$ can be written as
$\langle\vec r|\Psi\rangle=\big|\langle\vec r|\Psi\rangle\big|e^{i\phi(\vec r)}$, where $\phi(\vec r)$ is the phase, and
the expectation value of the superfluid circulation can be expressed as:
\begin{eqnarray}
  \nonumber \big\langle\hat\mathcal C\big\rangle_s &=& -i\hbar\big\langle\Psi\big|\oint_\Gamma\vec\nabla\cdot\textrm{d}\vec r\big|\Psi\big\rangle\\
  \nonumber                                        &=& -i\hbar\int_\Omega\oint_\Gamma\big\langle\Psi\big|\vec r\big\rangle\vec\nabla\big\langle\vec r\big|\Psi\big\rangle\cdot\textrm{d}\vec r\:\textrm{d}\Omega\\
  \nonumber                                        &=& -i\hbar\int_\Omega\oint_\Gamma\big|\big\langle\Psi\big|\vec r\big\rangle\big|\big(\vec\nabla\big|\big\langle\vec r\big|\Psi\big\rangle\big|\big)\cdot\textrm{d}\vec r\:\textrm{d}\Omega\\
  \label{AverageCirculation}                       && +\hbar\int_\Omega\oint_\Gamma\big\langle\Psi\big|\vec r\big\rangle\big(\vec\nabla\phi\big)\big\langle\vec r\big|\Psi\big\rangle\cdot\textrm{d}\vec r\:\textrm{d}\Omega
\end{eqnarray}
Because $\vec{\mathcal P}$ is Hermitian, only the real part in (\ref{AverageCirculation}) can be non-zero. Thus, the circulation
depends only on the phase gradient of the wave function and takes the form:
\begin{equation}
  \label{AverageCirculation2}\big\langle\hat\mathcal C\big\rangle_s=\hbar\big\langle\oint_\Gamma\vec\nabla\phi\cdot\textrm{d}\vec r\big\rangle_s
\end{equation}
Since the circulation of the phase gradient is along a closed loop, the total variation of the phase must vanish, unless the phase
runs over $n$ entire periods of $2\pi$. As a result, the circulation of the superfluid is quantized and can take only the values
$\big\langle\hat\mathcal C\big\rangle_s=2\pi n\hbar$. We note that, from Eq.~(\ref{AverageCirculation2}), the existence of a non-zero circulation
must be associated with a phase coherence.
Expressing the circulation in terms of the velocity $\vec v=\vec{\mathcal P}/m$, with $m$ the mass of one atom, we find the velocity quantization condition that was first proposed by
Onsager\cite{Onsager},
\begin{equation}
  \label{VelocityQuantization} \oint_\Gamma\vec v_s\cdot\textrm{d}\vec r=n\kappa_o,
\end{equation}
where $\vec v_s=\langle\vec v\rangle_s$ and $\kappa_o=\frac{2\pi\hbar}{m}$ is the flux quantum. If the integration loop does
not enclose a ``hole" (a physical hole like in the torus under consideration, or a vortex), then the path can be shrunk
continuously to a point where the circulation vanishes, corresponding to $n=0$. Thus, the only possibility for the circulation
to be non-zero is that the loop encloses at least one hole. For a loop that does not include
any hole, the application of Stokes' theorem implies that the superflow is irrotational:
\begin{equation}
  \label{Irrotational} \vec\nabla\times\vec v_s=0
\end{equation}
We can now make precise what we meant by \textit{low} and \textit{high} velocities. When the
initial velocity corresponds to less than half a flux quantum (low velocity), the superfluid seeks the nearest velocity satisfying (\ref{VelocityQuantization})
and comes to rest, thus excluding all flux (Meissner effect). When the inital velocity corresponds to more than half a flux quantum (high velocity), the superfluid
seeks the nearest velocity satisfying (\ref{VelocityQuantization}) and settles in a persistent dissipationless flow.

The discussion above suggests that transitions between states with different quantum numbers $n$ can be suspected to be
associated with vortex formations. The details behind those transitions have been studied numerically for the
case of the two-dimensional $XY$ model\cite{Batrouni}.

\section{Thought experiment and definitions of the normal and superfluid densities}
\subsection{Idealization}
\begin{figure}[h]
  \centerline{\includegraphics[width=0.45\textwidth]{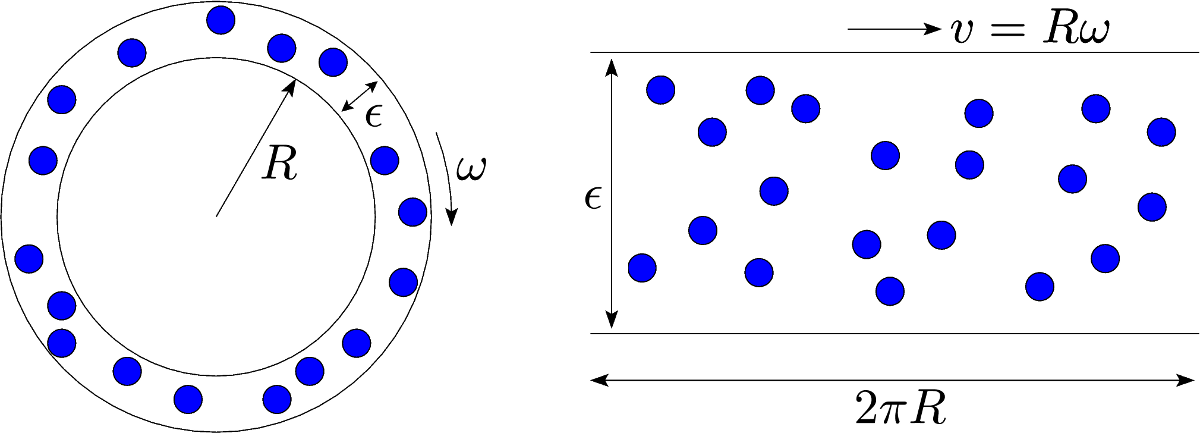}}
  \caption
    {
      (Color online) A fluid is enclosed between two hypercylinders of radii $R$ and $R+\epsilon$ rotating at angular
      frequency $\omega$ (left). In the limit $R\gg\epsilon$, the system becomes equivalent to one with a fluid enclosed between two hyperplanes
      moving at velocity $v=R\omega$, and with a periodicity of $2\pi R$ in the direction of $\vec v$ (right).
    }
  \label{ThoughtExperiment}
\end{figure}

We consider here a thought experiment that idealizes the above real experiment made
at low angular frequency. In this experiment, a fluid is enclosed between two \mbox{$d$-dimensional}
hypercylinders of radii $R$ and $R+\epsilon$ and infinite mass, rotating with angular frequency $\omega$ (Fig.~\ref{ThoughtExperiment}, left). We denote by $\mathcal F$ the frame attached
to the lab, and by $\mathcal F^\prime$ the frame attached to the moving walls. In the limit $R\gg\epsilon$, the system becomes equivalent to a fluid enclosed between two
hyperplanes of infinite mass separated by a distance $\epsilon$ and moving at constant velocity $v=R\omega$ with respect to $\mathcal F$, with
a periodicity of $2\pi R$ in the direction of $\vec v$ (Fig.~\ref{ThoughtExperiment}, right). The outcome of this thought experiment is that, at temperature below $T_c$,
the superfluid comes to rest with respect to the lab frame, while the normal fluid remains at rest in the frame of the moving walls.
Because of the infinite mass of the walls, their velocity with respect to $\mathcal F$ does not increase.

This ``thought outcome" does not contradict Galileo's principle of Relativity. It could be argued that
there is no reason for the superfluid to choose to come to rest with respect to $\mathcal F$ instead of any other inertial frame.
This apparent paradox is resolved by realizing that our description of the system enclosed
between two moving hyperplanes is just a limit of the system enclosed between two hypercylinders. Therefore, the frames $\mathcal F$
and $\mathcal F^\prime$ are not equivalent, $\mathcal F^\prime$ is actually rotating and thus accelerated, and coming to rest with respect to $\mathcal F$ is the only way for the superfluid
to avoid being accelerated.

\subsection{Isotropic case}
For a system described by an isotropic Hamiltonian,
this thought experiment can be directly generalized to systems with periodicity in all directions: We consider an orthonormal basis $\mathcal B=\big\lbrace\hat r_1,\cdots,\hat r_d\big\rbrace$
and a system with periodicity $L_j$ along each direction $\hat r_j$ with walls moving at velocity $\vec v=(v_1,\cdots,v_d)$ with
respect to the lab frame $\mathcal F$. As the temperature is lowered below $T_c$, the superfluid is observed to come to rest
with respect to $\mathcal F$, while the normal fluid remains at rest with respect to~$\mathcal F'$.
Because of the isotropy of the system, this observation must be independent of the direction of $\vec v$.
As a result, the normal density can be represented by a scalar $\rho_n$ and defined as the ratio $\rho_n=M_n/\Omega$, where
$M_n$ is the mass of the fluid that comes to rest with respect to the
frame of the moving walls $\mathcal F'$, and $\Omega$ is the volume of the system. In a similar way, the superfluid density can be represented by a
scalar $\rho_s$ and defined as the ratio $\rho_s=M_s/\Omega$, where $M_s$ is the mass of the fluid that comes to rest with respect to the lab frame $\mathcal F$.
In the following, we also consider the total mass of the fluid, $M=M_s+M_n$, and the total density, $\rho=\rho_s+\rho_n$.

\subsection{Anisotropic case}
For a system described by an anisotropic Hamiltonian (such as a crystal), the normal and superfluid currents are not necessarily parallel to the
velocity $\vec v$ of the walls This means that the normal and superfluid densities are
second-order tensors, $\rho_n^{\zeta\xi}$ and $\rho_s^{\zeta\xi}$. Using Einstein's summation convention, the components of
the normal and superfluid current densities $\vec j_n$ and $\vec j_s$ take the general form
\begin{eqnarray}
  \label{NormalCurrent}     && j_n^\zeta=\rho_n^{\zeta\xi}v_n^\xi,\\
  \label{SuperfluidCurrent} && j_s^\zeta=\rho_s^{\zeta\xi}v_s^\xi,
\end{eqnarray}
where $\vec v_n$ and $\vec v_s$ are the normal and superfluid velocities, with $\vec v_n=\vec v$ and $\vec v_s=0$ in $\mathcal F$,
and $\vec v_n=0$ and $\vec v_s=-\vec v$ in~$\mathcal F'$.

\section{\label{Section3} Calculation of the normal and superfluid densities in continuous space}
\subsection{Second quantization preliminaries and notations}
We give here a brief reminder of second quantization that is mainly meant to introduce the notation that we use all along this paper.
In second quantization, any operator can be expressed as a functional of the creation and annihilation field operators
$\hat\psi^\dagger(\vec r)$ and $\hat\psi(\vec r)$, with $\vec r=(r_1,\cdots,r_d)$, which satisfy the relations
\begin{eqnarray}
  \label{FieldCommutation1} && \big[\hat\psi(\vec r),\hat\psi(\vec r^{\:\prime})\big]_\zeta=\big[\hat\psi^\dagger(\vec r),\hat\psi^\dagger(\vec r^{\:\prime})\big]_\zeta=0,\\
  \label{FieldCommutation2} && \big[\hat\psi(\vec r),\hat\psi^\dagger(\vec r^{\:\prime})\big]_\zeta=\delta(\vec r^{\:\prime}-\vec r),
\end{eqnarray}
where $[A,B]_\zeta\stackrel{\textrm{\tiny def}}{=}AB-\zeta BA$ with $\zeta=1$ for bosons and $\zeta=-1$ for fermions, and $\delta(\vec r^{\:\prime}-\vec r)$ is the $d$-dimensional Dirac distribution.
The number operator $\hat\mathcal N$, the position operator $\vec{\mathcal R}$, and the momentum operator $\vec{\mathcal P}$
take the forms
\begin{eqnarray}
  \label{NumberOperator} \hat\mathcal N&=&\int_\Omega\hat\psi^\dagger(\vec r)\hat\psi(\vec r)\textrm{d}\Omega,\\
  \label{PositionOperator} \vec{\mathcal R}&=&\int_\Omega\vec r\:\hat\psi^\dagger(\vec r)\hat\psi(\vec r)\textrm{d}\Omega,\\
  \label{MomentumOperator} \vec{\mathcal P}&=&-i\hbar\int_\Omega\hat\psi^\dagger(\vec r)\vec\nabla_{\vec r}\hat\psi(\vec r)\textrm{d}\Omega,
\end{eqnarray}
with $\textrm{d}\Omega=\textrm{d}r_1\cdots \textrm{d}r_d$,
and satisfy the commutation relations:
\begin{eqnarray}
  \label{CanonicalCommutation} && \big[\mathcal R_\mu,\mathcal P_\nu\big]=i\hbar\delta_{\mu\nu}\hat\mathcal N\\
                               && \big[\vec{\mathcal R},\hat\mathcal N\big]=[\vec{\mathcal P},\hat\mathcal N\big]=0
\end{eqnarray}
From (\ref{FieldCommutation1}), (\ref{FieldCommutation2}), and (\ref{PositionOperator}), we can derive additional commulation
relations:
\begin{eqnarray}
  \label{PositionCreation}     && \big[\vec{\mathcal R},\hat\psi^\dagger(\vec r)]=\vec r\:\hat\psi^\dagger(\vec r)\\
  \label{PositionAnnihilation} && \big[\vec{\mathcal R},\hat\psi(\vec r)]=-\vec r\:\hat\psi(\vec r)
\end{eqnarray}

\subsection{Continuous isotropic case}
We consider a $d$-dimensional system of identical particles of mass $m$ in a $L_1\times\cdots\times L_d$ box with periodic boundary conditions that is moving at
low velocity $\vec v$ with respect to the lab frame $\mathcal F$.
In the frame $\mathcal F'$ of the moving walls the system is at rest. Thus, in this frame, the Hamiltonian is independent of the velocity $\vec v$,
and is a functional $\Phi$ of the creation and annihilation field operators:
\begin{equation}
  \label{Hamiltonian1} \hat\mathcal H_o=\Phi[\hat\psi^\dagger(\vec r),\hat\psi(\vec r)]
\end{equation}
Defining the partition function $\mathcal Z_o=\textrm{Tr }e^{-\beta\hat\mathcal H_o}$ where $\beta$ is the inverse temperature,
the average total momentum in $\mathcal F'$ is given by:
\begin{equation}
  \big\langle\vec{\mathcal P}\big\rangle_{\mathcal F'}=\frac{1}{\mathcal Z_o}\textrm{Tr }\vec{\mathcal P}e^{-\beta\hat\mathcal H_o}
\end{equation}
Since $\mathcal F$ moves at velocity $-\vec v$ with respect to $\mathcal F'$, the total momentum in the lab frame can be obtained from
the above expression by performing an inverse Galilean transformation with velocity $\vec v$.
For this purpose, it is useful to define the unitary operator:
\begin{equation}
  \label{GalileanOperator} \hat\mathcal U=e^{-i\frac{m}{\hbar}\vec v\cdot\vec{\mathcal R}}
\end{equation}
By using (\ref{CanonicalCommutation}), it is straightforward to check that:
\begin{eqnarray}
  && \hat\mathcal U^\dagger\vec{\mathcal P}\hat\mathcal U=\vec{\mathcal P}-m\vec v\hat\mathcal N\\
  && \hat\mathcal U^\dagger\vec{\mathcal R}\hat\mathcal U=\vec{\mathcal R}
\end{eqnarray}
Thus, $\hat\mathcal U$ is the operator that performs a Galilean transformation (at time $t=0$) with velocity~$\vec v$, and the
total momentum operator in $\mathcal F$ is given by the inverse transformation, $\hat\mathcal U\vec{\mathcal P}\hat\mathcal U^\dagger$.
Since the density matrix $e^{-\beta\hat\mathcal H_o}/\mathcal Z_o$ describes probabilities of states, it remains unchanged when
going to the lab frame. As a result, the average total momentum in $\mathcal F$ takes the form
\begin{eqnarray}
  \nonumber \big\langle\vec{\mathcal P}\big\rangle_{\mathcal F} &=& \frac{1}{\mathcal Z_o}\textrm{Tr }\hat\mathcal U\vec{\mathcal P}\hat\mathcal U^\dagger e^{-\beta\hat\mathcal H_o}\\
  \nonumber                                                     &=& \frac{1}{\mathcal Z_o}\textrm{Tr }\vec{\mathcal P} e^{-\beta\hat\mathcal U^\dagger\hat\mathcal H_o\hat\mathcal U}\\
  \label{GalileanTransformation}                                &=& \frac{1}{\mathcal Z_o}\textrm{Tr }\vec{\mathcal P} e^{-\beta\hat\mathcal H_{\vec v}},
\end{eqnarray}
where we have used the invariance of the trace under cyclic permutations, and we have defined:
\begin{equation}
  \label{HamiltonianTransform} \hat\mathcal H_{\vec v}=\hat\mathcal U^\dagger\hat\mathcal H_o\hat\mathcal U
\end{equation}
From the correspondence principle, the classical momentum of the fluid must be equal to the quantum average of the momentum
operator $\vec{\mathcal P}$. Since in $\mathcal F$ only the normal fluid with mass $M_n=\rho_n\Omega$ spins, we have:
\begin{equation}
  \label{CorrespondencePrinciple} \rho_n\Omega\vec v=\frac{1}{\mathcal Z_o}\textrm{Tr }\vec{\mathcal P}e^{-\beta\hat\mathcal H_{\vec v}}
\end{equation}
Calculating the divergence of (\ref{CorrespondencePrinciple}) with respect to $\vec v$, we get:
\begin{eqnarray}
  \nonumber \vec\nabla_{\vec v}\cdot(\rho_n\Omega\vec v) &=& \Omega\big(\vec\nabla_{\vec v}\rho_n\big)\cdot\vec v+\rho_n\Omega d\\
  \label{Expression}                                     && \!\!\!\!\!\!\!\!\!\!\!\!\!\!\!\!\!\!\!\!\!\!\!\!\!\!\!\!\!\!\!\!=-\frac{1}{\mathcal Z_o}\textrm{Tr }\vec{\mathcal P}\cdot\!\!\int_0^\beta\!\!\!\! e^{-(\beta-\tau)\hat\mathcal H_{\vec v}}\vec\nabla_{\vec v}\hat\mathcal H_{\vec v} e^{-\tau\hat\mathcal H_{\vec v}}\textrm{d}\tau
\end{eqnarray}
From (\ref{GalileanOperator}) and (\ref{HamiltonianTransform}), the gradient of $\hat\mathcal H_{\vec v}$ with respect to $\vec v$ can be expressed as:
\begin{equation}
  \label{GradientHamiltonian} \vec\nabla_{\vec v}\hat\mathcal H_{\vec v}=i\frac{m}{\hbar}\hat\mathcal U^\dagger\big[\vec{\mathcal R},\hat\mathcal H_o\big]\hat\mathcal U
\end{equation}
Injecting (\ref{GradientHamiltonian}) in (\ref{Expression}) and taking the limit $\vec v\to 0$, 
we get the expression of the normal density:
\begin{equation}
  \label{NormalDensity} \rho_n=-i\frac{m}{\hbar\Omega d}\Big\langle\vec{\mathcal P}\cdot\int_0^\beta e^{\tau\hat\mathcal H_o}\big[\vec{\mathcal R},\hat\mathcal H_o\big]e^{-\tau\hat\mathcal H_o}\textrm{d}\tau\Big\rangle
\end{equation}
From the relation $\rho_n+\rho_s=\rho$, we deduce the expression of the superfluid density:
\begin{equation}
  \label{SuperfluidDensity} \rho_s=\rho+i\frac{m}{\hbar\Omega d}\Big\langle\vec{\mathcal P}\cdot\int_0^\beta e^{\tau\hat\mathcal H_o}\big[\vec{\mathcal R},\hat\mathcal H_o\big]e^{-\tau\hat\mathcal H_o}\textrm{d}\tau\Big\rangle
\end{equation}
The above expressions of $\rho_n$ and $\rho_s$ can be evaluated with path-integrals methods, such as the Stochastic Green
Function (SGF) algorithm\cite{SGF,DirectedSGF} (see paragraphs~\ref{Example} and~\ref{Example2} for concrete examples).

\subsection{Continuous anisotropic case}
All equations in the previous subsection can be easily generalized to anisotropic Hamiltonians. By definition, the total momentum
in $\mathcal F$ is obtained by integrating the normal current density $\vec j_n$ over the volume $\Omega$. Assuming
for the sake of simplicity a uniform current density, we have $\big\langle\vec P\big\rangle_{\mathcal F}=\vec j_n\Omega$. Using
(\ref{NormalCurrent}) in $\mathcal F$, the expression of the correspondence principle (\ref{CorrespondencePrinciple}) is generalized as:
\begin{equation}
  \label{GalileanTransformation2}\rho_n^{\zeta\xi}v_\xi\Omega=\frac{1}{\mathcal Z_o}\textrm{Tr }\mathcal P_\zeta e^{-\beta\hat\mathcal H_{\vec v}}
\end{equation}
Calculating the derivative with respect to $v_\xi$, and taking the limit $\vec v\to 0$, the normal density tensor takes the form:
\begin{equation}
  \label{NormalDensity2} \rho_n^{\zeta\xi}=-i\frac{m}{\hbar\Omega}\Big\langle\mathcal P_\zeta\int_0^\beta e^{\tau\hat\mathcal H_o}\big[\mathcal R_\xi,\hat\mathcal H_o\big]e^{-\tau\hat\mathcal H_o}\textrm{d}\tau\Big\rangle
\end{equation}
In order to obtain the superfluid density tensor, it is convenient to consider the total momentum in $\mathcal F'$. By definition,
we have $\big\langle\vec P\big\rangle_{\mathcal F'}=\vec j_s\Omega$. Using (\ref{SuperfluidCurrent}) in $\mathcal F'$, the
correspondence principle implies:
\begin{eqnarray}
  \nonumber -\rho_s^{\zeta\xi}v_\xi\Omega &=& \frac{1}{\mathcal Z_o}\textrm{Tr }\mathcal P_\zeta e^{-\beta\hat\mathcal H_o}\\
  \nonumber                               &=& \frac{1}{\mathcal Z_o}\textrm{Tr }\hat\mathcal U^\dagger\mathcal P_\zeta\hat\mathcal U e^{-\beta\hat\mathcal U^\dagger\hat\mathcal H_o\hat\mathcal U}\\
                                          &=& \rho_n^{\zeta\xi}v_\xi\Omega-mv_\zeta\big\langle\hat\mathcal N\big\rangle
\end{eqnarray}
Calculating the derivative with respect to $v_\xi$, and taking the limit $\vec v\to 0$, the superfluid density tensor is obtained
from the normal density tensor (\ref{NormalDensity2}) as:
\begin{equation}
  \label{SuperfluidDensityTensor} \rho_s^{\zeta\xi}=\rho\delta_{\zeta\xi}-\rho_n^{\zeta\xi}
\end{equation}

\subsection{Relationship between the superfluid density and the free energy}
For the simplicity of the following discussion, we consider here only the isotropic case, the generalization to the anisotropic
case being straightforward.
There exists a particular class of Hamiltonians for which the superfluid density can be directly related to the Laplacian $\Delta_{\vec v}$
of the free energy associated to the Hamiltonian $\hat\mathcal H_{\vec v}$,
\begin{equation}
  \label{FreeEnergy}F_{\vec v}=-\frac{1}{\beta}\ln\mathcal Z_{\vec v},
\end{equation}
with $\mathcal Z_{\vec v}=\textrm{Tr }e^{-\beta\hat\mathcal H_{\vec v}}$.
This class is defined by Hamiltonians $\hat\mathcal H_o$ for which the commutator with the position operator satisfies:
\begin{equation}
  \label{ParticularClass} \big[\vec{\mathcal R},\hat\mathcal H_o\big]=i\frac{\hbar}{m}\vec{\mathcal P}
\end{equation}
The most common example of Hamiltonian that belongs to this class is given by $\hat\mathcal H_o=\hat\mathcal T+\hat\mathcal V$,
where $\hat\mathcal V$ is a potential that satisfies $\big[\vec{\mathcal R},\hat\mathcal V\big]=0$
and $\hat\mathcal T$ is the kinetic energy:
\begin{equation}
  \label{KineticEnergy}\hat\mathcal T=-\frac{\hbar^2}{2m}\int_\Omega\hat\psi^\dagger(\vec r)\Delta_{\vec r}\hat\psi(\vec r)\textrm{d}\Omega
\end{equation}
For any Hamiltonian that satisfies (\ref{ParticularClass}), the gradient (\ref{GradientHamiltonian}) takes the form
\begin{equation}
  \vec\nabla_{\vec v}\hat\mathcal H_{\vec v}=-\vec{\mathcal P}+m\vec v\hat\mathcal N,
\end{equation}
from which $\vec{\mathcal P}$ can be extracted and injected into (\ref{CorrespondencePrinciple}), leading to:
\begin{eqnarray}
  \nonumber\rho_n\Omega\vec v &=& -\frac{1}{\mathcal Z_o}\textrm{Tr }\big(\vec\nabla_{\vec v}\hat\mathcal H_{\vec v}-m\vec v\hat\mathcal N\big)e^{-\beta\hat\mathcal H_{\vec v}}\\
                              &=& -\frac{\mathcal Z_{\vec v}}{\mathcal Z_o}\vec\nabla_{\vec v}F_{\vec v}+\frac{m\vec v}{\mathcal Z_o}\textrm{Tr }\hat\mathcal Ne^{-\beta\hat\mathcal H_{\vec v}}
\end{eqnarray}
From the divergence of the above expression in the limit $\vec v\to 0$, we get for the superfluid density the expression:
\begin{equation}
  \label{RhoSFreeEnergy} \rho_s=\lim_{\vec v\to 0}\frac{1}{\Omega d}\Delta_{\vec v} F_{\vec v}
\end{equation}
At this point, it is useful to determine how the creation and annihilation field operators transform under $\hat\mathcal U$. Using (\ref{PositionCreation})
and (\ref{PositionAnnihilation}), we find:
\begin{eqnarray}
  \label{CreationPhaseBoost}     &&\hat\mathcal U^\dagger\hat\psi^\dagger(\vec r)\hat\mathcal U=\hat\psi^\dagger(\vec r)e^{i\frac{m}{\hbar}\vec v\cdot\vec r}\\
  \label{AnnihilationPhaseBoost} &&\hat\mathcal U^\dagger\hat\psi(\vec r)\hat\mathcal U=\hat\psi(\vec r)e^{-i\frac{m}{\hbar}\vec v\cdot\vec r}
\end{eqnarray}
As a result, performing a Galilean transformation with velocity $\vec v$ is equivalent to applying a phase boost $\varphi(\vec r)=\frac{m}{\hbar}\vec v\cdot\vec r$
to the creation and annihilation field operators.
This allows us to relate the superfluid density to the response of the free energy to a boundary phase-twist\cite{Fisher}.
For this, consider a vector $\vec L=n_1L_1\hat r_1+\cdots+n_dL_d\hat r_d$ where $n_1,\cdots,n_d$ are integers.
The phase-twist at the tips of the vector $\vec L$ that results from the velocity $\vec v$ is $\phi=\frac{m}{\hbar}\vec v\cdot\vec L$.
This allows us to rewrite (\ref{RhoSFreeEnergy}) as:
\begin{equation}
  \label{RhoSFreeEnergy2} \rho_s=\lim_{\phi\to 0}\frac{m^2\vec L^2}{\hbar^2\Omega d}\frac{\partial^2F_\phi}{\partial\phi^2}
\end{equation}
It is clear that the free energy cannot depend on the sign of the velocity or the phase-twist.
This implies:
\begin{equation}
  \lim_{\phi\to 0}\frac{\partial^2 F_\phi}{\partial\phi^2}=\lim_{\phi\to 0}\frac{1}{\phi}\frac{\partial F_\phi}{\partial\phi}
\end{equation}
Therefore, the superfluid density is directly related to the response of the free energy to a boundary phase-twist\cite{Fisher}:
\begin{equation}
  \label{RhoSFreeEnergy3} \rho_s=\lim_{\phi\to 0}\frac{m^2\vec L^2}{\hbar^2\Omega d\phi}\frac{\partial F_\phi}{\partial\phi}
\end{equation}
While equations (\ref{RhoSFreeEnergy}), (\ref{RhoSFreeEnergy2}), and (\ref{RhoSFreeEnergy3}) are well known, it is important
to keep in mind that they are valid only for Hamiltonians that satisfy Eq.~(\ref{ParticularClass}).

\subsection{Relationship between the superfluid density and the winding number}
As in the previous subsection, we assume here that the Hamiltonian satisfies the condition (\ref{ParticularClass}). In addition,
we add the constraint that the Hamiltonian conserves the number of particles, $\big[\hat\mathcal H_o,\hat\mathcal N\big]=0$.
Performing a Taylor expansion and introducing $n$ complete sets of states $\big|\Psi_k\big\rangle$ in the position occupation
number representation, the partition function $\mathcal Z_{\vec v}$
can be written as
\begin{equation}
  \label{PartitionFunction1} \mathcal Z_{\vec v}=\sum_{n\geq 0}\frac{(-\beta)^n}{n!}\!\!\!\!\!\!\sum_{\Psi_1\cdots\Psi_n}\prod_{k=1}^n\big\langle\Psi_{k+1}\big|\hat\mathcal H_{\vec v}\big|\Psi_k\big\rangle,
\end{equation}
with the convention $\big|\Psi_{n+1}\big\rangle=\big|\Psi_1\big\rangle$.
From (\ref{HamiltonianTransform}), we have
\begin{eqnarray}
  \nonumber \prod_{k=1}^n\big\langle\Psi_{k+1}\big|\hat\mathcal H_{\vec v}\big|\Psi_k\big\rangle &=& \prod_{k=1}^n\big\langle\Psi_{k+1}\big|\hat\mathcal H_o\big|\Psi_k\big\rangle\\
  &\times& e^{i\frac{m}{\hbar}\vec v\cdot\sum_j L_jW_j^{\Psi}\hat r_j},
\end{eqnarray}
where $W_j^{\Psi}$ counts the number of particles that cross the boundaries of the system in the direction $\hat r_j$ while evolving over
the sequence of states $\Psi_k$.
Therefore the partition function $\mathcal Z_v$ can be written as\cite{Batrouni2}
\begin{eqnarray}
  \nonumber\mathcal Z_v      &=& \sum_{n\geq 0}\sum_{\Psi_1\cdots\Psi_n}\!\!\!\!\!\!\!\!\!\!\!\!\!\!\underbrace{\frac{(-\beta)^n}{n!}\!\prod_{k=1}^n\big\langle\Psi_{k+1}\big|\hat\mathcal H_o\big|\Psi_k\big\rangle}_{\textrm{\tiny Boltzmann weight of a configuration }\Psi_1\cdots\Psi_n}\!\!\!\!\!\!\!\!\!\! e^{i\frac{m}{\hbar}\vec v\cdot\sum_j L_j W_j^\Psi\hat r_j}\\
  \label{PartitionWinding}   &=& \mathcal Z_o\big\langle e^{i\frac{m}{\hbar}\vec v\cdot\sum_j L_j\mathcal W_j\hat r_j}\big\rangle,
\end{eqnarray}
where $\mathcal W_j$ are the components of the winding number operator $\vec{\mathcal W}$
that take the eigenvalues $W_j^\Psi$ in a configuration of states $\Psi_1\cdots\Psi_n$.
Injecting (\ref{PartitionWinding}) into (\ref{FreeEnergy}), and using (\ref{RhoSFreeEnergy}),
the superfluid density becomes directly related to the fluctuations of the winding number:
\begin{equation}
  \label{SuperfluidWinding} \rho_s=\frac{m^2}{\hbar^2\beta\Omega d}\Big\langle\Big(\sum_j L_j\mathcal W_j\hat r_j\Big)^2\Big\rangle
\end{equation}
For a hypercubic system, $\Omega=L^d$, the above expression becomes\cite{PollockCeperley}:
\begin{equation}
  \label{PollockCeperleySuperfluid} \rho_s=\frac{m^2 L^{2-d}}{\hbar^2\beta d}\big\langle\vec{\mathcal W}^2\big\rangle
\end{equation}
Here too, it is important to keep in mind that while Eq.~(\ref{PollockCeperleySuperfluid}) is well known, it cannot be applied
to Hamiltonians that do not satisfy Eq.~(\ref{ParticularClass}). In addition, the conservation of the number of particles
is required for the winding number to be well defined.
For Hamiltonians that do not satisfy these conditions, only equations~(\ref{NormalDensity}) and~(\ref{SuperfluidDensity}) are valid.

\section{\label{Section4} Calculation of the normal and superfluid densities in discrete space}
Determining the expression of the superfluid density in discrete space is not as straightforward as it looks like. In particular,
simply replacing the continuous-space operators by their discrete-space equivalents into (\ref{SuperfluidDensity}) leads to inconsistencies.
The reason is that some of the usual commutation
rules between the operators $\hat\mathcal N$, $\vec{\mathcal R}$, $\vec{\mathcal P}$, and $\hat\mathcal T$ are no longer valid
when these operators are discretized. It is therefore necessary to proceed carefully
with the discretization of space.

\subsection{Discretization of space}
We start by noticing that (\ref{NumberOperator}) represents a dimensionless  quantity. This implies that the dimension of
the creation and annihilation field operators is the inverse squareroot of a $d$-dimensional volume:
\begin{equation}
  \big[\hat\psi^\dagger(\vec r)\big]=\big[\hat\psi(\vec r)\big]=\mathcal L^{-\frac{d}{2}}
\end{equation}
Performing for each components of $\vec r$ the change of variable $r_j=l_jy_j$, where $l_j$ is a positive parameter with the dimension of a length and $y_j$ is
the new dimensionless variable, we can define the dimensionless creation and annihilation operators
\begin{eqnarray}
  \label{DiscreteField1} && a^\dagger_{\vec y}=\sqrt{l_1\cdots l_d}\:\hat\psi^\dagger(l_1y_1,\cdots,l_dy_d),\\
  \label{DiscreteField2} && a_{\vec y}=\sqrt{l_1\cdots l_d}\:\hat\psi(l_1y_1,\cdots,l_dy_d),
\end{eqnarray}
which satisfy the same commulation relations as (\ref{FieldCommutation1}) and (\ref{FieldCommutation2}).
Using these dimensionless operators, Eq.~(\ref{NumberOperator}) can be rewritten as
\begin{equation}
  \hat\mathcal N=\int_{\Omega} a^\dagger_{\vec y}a_{\vec y}^{\phantom\dagger}\textrm{d}y_1\cdots\textrm{d}y_d,
\end{equation}
In the limit $l_j\to 0$ the integral becomes independent on the step size $\textrm{d}y_j$, which can be chosen to be unity. As
a result, the continuous integral can be replaced by a discrete sum over a lattice with constants $l_1,\cdots,l_d$.
The commutation relation between the annihilation and the creation operator becomes:
\begin{equation}
  \label{Commutator4}\big[a_{\vec y}^{\phantom\dagger},a_{\vec y'}^\dagger\big]_\zeta=\delta_{\vec y\vec y^{\:\prime}}
\end{equation}
Defining $\hat n_{\vec y}=a_{\vec y}^\dagger a_{\vec y}^{\phantom\dagger}$, the number operator takes the well-known form:
\begin{equation}
  \label{DiscreteN}\hat\mathcal N=\sum_{\vec y}\hat n_{\vec y}
\end{equation}
Applying the same discretization procedure to the position operator, we get
\begin{equation}
  \label{DiscreteR}\vec{\mathcal R}=\sum_{\vec y}\vec y_{\vec l}\:\hat n_{\vec y},
\end{equation}
with $\vec y_{\vec l}=(l_1y_1,\cdots,l_dy_d)$.
By using for the first-order derivative the symmetrical prescription
\begin{equation}
  \label{FirstPrescription}\frac{\partial}{\partial y_j}a_{\vec y}=\frac{1}{2}\big(a_{\vec y+\hat j}-a_{\vec y-\hat j}\big),
\end{equation}
the discrete momentum operator takes the form
\begin{equation}
  \label{DiscreteP}\vec{\mathcal P}=-i\sqrt{mt_j/2}\sum_{\vec y}\big(a_{\vec y}^\dagger a_{\vec y+\hat j}-H.c.\big)\hat j,
\end{equation}
where the sum over $j$ is implicit and we have defined $t_j=\hbar^2/2ml_j^2$.
The above quantity is proportional to what is commonly known as the \textit{current density operator}\cite{Scalapino}. In this manuscript
we prefer to call it discrete momentum, as it converges to the continuous momentum when the lattice constant goes to zero.
It is important to emphasize here that (\ref{DiscreteP}) represents a discretization of the real momentum of the system, and that it should not be
confused with the crystal quasi-momentum. The importance of this distinction is made clear below.
Using for the second-order derivative the symmetrical prescription
\begin{eqnarray}
  \nonumber \frac{\partial^2}{\partial y_{\mu}\partial y_{\nu}}a_{\vec y}&=&\frac{1}{2}\big(a_{\vec y+\hat\mu}+a_{\vec y-\hat\mu}+a_{\vec y+\hat\nu}+a_{\vec y-\hat\nu}\\
  \label{SecondPrescription}                                             &&-a_{\vec y+\hat\mu-\hat\nu}-a_{\vec y-\hat\mu+\hat\nu}-2a_{\vec y}\big),
\end{eqnarray}
the discrete kinetic operator takes the form $\hat\mathcal T=\sum_j\hat\mathcal T_j$, where $\hat\mathcal T_j$ is given by:
\begin{equation}
  \label{DiscreteT}\hat\mathcal T_j=-t_j\sum_{\vec y}\big(a_{\vec y}^\dagger a_{\vec y+\hat j}+H.c.\big)+2t_j\hat\mathcal N
\end{equation}
Note that the second term in (\ref{DiscreteT}) is usually dismissed because it only gives rise to a shift of the chemical potential
and does not change the physics. In our case, we explicitly take it into account in order to ease the connection with the continuous
case.
With these discrete operators, it is easy to check that the commutator (\ref{CanonicalCommutation}) becomes:
\begin{equation}
  \label{DiscreteCanonicalCommutation}\big[\mathcal R_\mu,\mathcal P_\nu\big]=i\hbar\delta_{\mu\nu}\Big(\hat\mathcal N-\frac{1}{2t_\mu}\hat\mathcal T_\mu\Big)
\end{equation}
As a result, as opposed to the continuous case, the discrete position operator $\vec{\mathcal R}$ is not the generator of infinitesimal
translations in real momentum space. It actually translates the quasi-momentum only. Therefore the unitary operator $\hat\mathcal U=e^{-i\frac{m}{\hbar}\vec v\cdot\vec{\mathcal R}}$
is no longer the operator that performs a Galilean transformation with velocity $\vec v$. This implies that Eq.~(\ref{GalileanTransformation})
and (\ref{GalileanTransformation2}) are not applicable in the discrete case and need to be modified. To this end, it is
useful to determine how the real momentum $\vec{\mathcal P}$ transforms under $\hat\mathcal U$ at first order in $\vec v$.
Using (\ref{DiscreteCanonicalCommutation}) we find:
\begin{equation}
  \label{DiscreteUnitaryTransformation}\hat\mathcal U\vec{\mathcal P}\hat\mathcal U^\dagger=\vec{\mathcal P}+m\vec v\hat\mathcal N-\frac{m}{2}\sum_j \frac{v_j}{t_j}\hat j\hat\mathcal T_j+\vec{\mathcal O}(\vec v^2)
\end{equation}
This implies that new terms proportional to $\big\langle\hat\mathcal T_j\big\rangle$ must be introduced when discretizing (\ref{GalileanTransformation})
or (\ref{GalileanTransformation2}).

\subsection{Discrete isotropic case}
Using (\ref{DiscreteUnitaryTransformation}) with $t_j=t$, (\ref{GalileanTransformation}) becomes:
\begin{equation}
  \label{DiscreteCorrespondencePrinciple}\nonumber\big\langle\vec{\mathcal P}\big\rangle_{\mathcal F}=\frac{1}{\mathcal Z_o}\textrm{Tr }\vec{\mathcal P} e^{-\beta\hat\mathcal H_{\vec v}}+\frac{m}{2t}\sum_j v_j\hat j\big\langle\hat\mathcal T_j\big\rangle+\vec{\mathcal O}(\vec v^2)
\end{equation}
As before, the correspondence principle requires the above quantum average of the momentum to be equal to the
classical momentum, which in $\mathcal F$ is due to the normal fluid only, $\rho_n\Omega\vec v$. Calculating the divergence
of this equality and taking the limit $\vec v\to 0$, we deduce the expression of the normal density:
\begin{equation}
  \label{DiscreteNormalDensity} \rho_n=\frac{m}{2td\Omega}\big\langle\hat\mathcal T\big\rangle-i\frac{m}{\hbar\Omega d}\Big\langle\vec{\mathcal P}\cdot\int_0^\beta e^{\tau\hat\mathcal H_o}\big[\vec{\mathcal R},\hat\mathcal H_o\big]e^{-\tau\hat\mathcal H_o}\textrm{d}\tau\Big\rangle
\end{equation}
As a result, the expression of the superfluid density is:
\begin{eqnarray}
  \nonumber\rho_s                   &=&\rho-\frac{m}{2td\Omega}\big\langle\hat\mathcal T\big\rangle\\
  \label{DiscreteSuperfluidDensity} && +i\frac{m}{\hbar\Omega d}\Big\langle\vec{\mathcal P}\cdot\int_0^\beta e^{\tau\hat\mathcal H_o}\big[\vec{\mathcal R},\hat\mathcal H_o\big]e^{-\tau\hat\mathcal H_o}\textrm{d}\tau\Big\rangle
\end{eqnarray}
Comparing (\ref{SuperfluidDensity}) and (\ref{DiscreteSuperfluidDensity}), we see that not only have the continuous-space operators
been replaced by their discrete-space equivalents, but a new term proportional to the kinetic energy also appeared. It can be checked that
(\ref{DiscreteNormalDensity}) and (\ref{DiscreteSuperfluidDensity}) converge respectively to (\ref{NormalDensity}) and (\ref{SuperfluidDensity}) in the limit $l\to 0$.
The consistency of Eq.(\ref{DiscreteSuperfluidDensity}) can also be checked by verifying that it reduces to Eq.(13) of
Ref.\cite{Scalapino} when applied to the particular case discussed there.

\subsection{Discrete anisotropic case}
Using (\ref{DiscreteUnitaryTransformation}), (\ref{GalileanTransformation2}) becomes:
\begin{equation}
  \rho_n^{\zeta\xi}v_\xi\Omega=\frac{1}{\mathcal Z_o}\textrm{Tr }\mathcal P_\zeta e^{-\beta\hat\mathcal H_{\vec v}}+\frac{mv_\zeta}{2t_\zeta}\big\langle\hat\mathcal T_\zeta\big\rangle+\mathcal O(\vec v^2)
\end{equation}
Calculating the derivative with respect to $v_\xi$ in the limit $\vec v\to 0$, we get the normal density tensor:
\begin{eqnarray}
  \nonumber \rho_n^{\zeta\xi} &=& \frac{m}{2\Omega t_\zeta}\delta_{\zeta\xi}\big\langle\hat\mathcal T_\zeta\big\rangle\\
                              && -i\frac{m}{\hbar\Omega}\Big\langle\mathcal P_\zeta\int_0^\beta e^{\tau\hat\mathcal H_o}\big[\mathcal R_\xi,\hat\mathcal H_o\big]e^{-\tau\hat\mathcal H_o}\textrm{d}\tau\Big\rangle
\end{eqnarray}
As before, the superfluid density tensor is obtained as a function of the normal density tensor (\ref{SuperfluidDensityTensor}).

\subsection{The superfluid density as a function of the free energy and the winding number}
For simplicity, we consider in the remaining of this section only the isotropic case, the generalization to the anisotropic case being straightforward.
As for the continuous-space case, the superfluid density can be related to the free energy if the Hamiltonian satisfies $\big[\vec{\mathcal R},\hat\mathcal H_o\big]=i\frac{\hbar}{m}\vec{\mathcal P}$.
In this case, the gradient $\vec\nabla\hat\mathcal H_{\vec v}=-\hat\mathcal U^\dagger\vec{\mathcal P}\hat\mathcal U$ is given by the opposite of the inverse
transfomation of (\ref{DiscreteUnitaryTransformation}):
\begin{equation}
  \label{DiscreteGradient}\vec\nabla\hat\mathcal H_{\vec v}=-\vec{\mathcal P}+m\vec v\hat\mathcal N-\frac{m}{2t}\sum_j v_j\hat j\hat\mathcal T_j+\vec{\mathcal O}(\vec v^2)
\end{equation}
Extracting $\vec{\mathcal P}$ from the above expression and injecting it into (\ref{DiscreteCorrespondencePrinciple}), we get:
\begin{eqnarray}
  \nonumber \big\langle\vec{\mathcal P}\big\rangle_{\mathcal F} &=& -\frac{\mathcal Z_{\vec v}}{\mathcal Z_o}\vec\nabla F_{\vec v}+m\vec v\frac{1}{\mathcal Z_o}\textrm{Tr }\hat\mathcal N e^{-\beta\hat\mathcal H_{\vec v}}\\
  \nonumber                                                     && -\frac{m}{2t}\sum_j v_j\hat j\frac{1}{\mathcal Z_o}\textrm{Tr }\hat\mathcal T_j e^{-\beta\hat\mathcal H_{\vec v}}\\
  \label{DiscreteCorrespondencePrinciple2}                      && +\frac{m}{2t}\sum_j v_j\hat j\big\langle\hat\mathcal T_j\big\rangle+\vec{\mathcal O}(\vec v^2)
\end{eqnarray}
Calculating the divergence of (\ref{DiscreteCorrespondencePrinciple2}) and taking the limit $\vec v\to 0$, the terms in $\hat\mathcal T_j$
cancel out. Thus, the expression of $\rho_s$ as a function of the free energy in discrete space is the same
as in continuous space  (\ref{RhoSFreeEnergy}), and so are the expressions of $\rho_s$ as the response of the free energy to a phase boost (\ref{RhoSFreeEnergy3}),
and as a function of the winding number (\ref{SuperfluidWinding}).

\subsection{Dimensionless superfluid density}
For lattice systems, it is common to work with the dimensionless superfluid density $\tilde\rho_s$, defined as the superfluid
fraction $\eta_s=\rho_s/\rho$ times the dimensionless density $\tilde\rho=\langle\hat\mathcal N\rangle/S_{\textrm{\tiny tot}}$,
where $S_{\textrm{\tiny tot}}=S_1\times\cdots\times S_d$ is the
total number of lattice sites. This can be written as:
\begin{equation}
  \label{DimensionlessRhoS} \tilde\rho_s=\frac{\rho_s\Omega}{mS_{\textrm{\tiny tot}}}
\end{equation}
Injecting (\ref{DiscreteSuperfluidDensity}) into (\ref{DimensionlessRhoS}), the dimensionless superfluid density takes the
general form:
\begin{eqnarray}
  \nonumber           \tilde\rho_s &=& \tilde\rho-\frac{1}{2tdS_{\textrm{\tiny tot}}}\big\langle\hat\mathcal T\big\rangle\\
  \label{DimensionlessGeneralRhoS} && +i\frac{1}{\hbar dS_{\textrm{\tiny tot}}}\Big\langle\vec{\mathcal P}\cdot\int_0^\beta e^{\tau\hat\mathcal H_o}\big[\vec{\mathcal R},\hat\mathcal H_o\big]e^{-\tau\hat\mathcal H_o}\textrm{d}\tau\Big\rangle
\end{eqnarray}

\section{Lattices with non-orthonormal primitive vectors}
In order to obtain the correct expressions of the superfluid density in lattices with non-cubic primitive cells, it is necessary
to perform a careful change of basis when discretizing space.

\subsection{Change of basis}
We consider an orthonormal basis, $\mathcal B_r=\lbrace\hat r_1,\cdots,\hat r_d\rbrace$, and a
transformation $\mathcal A$ that changes $\mathcal B_r$ into a general basis, $\mathcal B_q=\lbrace\vec q_1,\cdots,\vec q_d\rbrace$.
We denote by $A$ the matrix representation of $\mathcal A$ in the basis $\mathcal B_r$. Position vectors are contravariant,
thus their coordinates $(q^1,\cdots,q^d)$ in $\mathcal B_q$ are obtained from their coordinates $(r^1,\cdots,r^d)$ in $\mathcal B_r$
by the inverse transformation,
\begin{equation}
  \label{Contravariant} q^i=A^{-1}_{ij}r^j,
\end{equation}
where we have used Einstein's summation convention.
On the contrary, the derivatives with respect to the coordinates are covariant,
\begin{equation}
  \frac{\partial}{\partial q_j}=A_{ij}\frac{\partial}{\partial r_i},
\end{equation}
thus the Laplacian in the basis $\mathcal B_q$ takes the form:
\begin{equation}
  \Delta_{\vec q}=A^{-1}_{ik}A^{-1}_{jk}\frac{\partial^2}{\partial q_i\partial q_j}
\end{equation}
Defining the metric tensor $g_{\mu\nu}=\vec q_\mu\cdot\vec q_\nu$,
the dot-product in the basis $\mathcal B_q$ of two vectors $\vec u=(u^1,\cdots,u^d)$ and $\vec v=(v^1,\cdots,v^d)$ takes the form:
\begin{equation}
  \label{DotProduct} \vec u\cdot\vec v=g_{\mu\nu}u^\mu v^\nu
\end{equation}

\subsection{Discretization of space in non-orthonormal coordinates}
For the sake of simplicity, we assume in the remainder of this section that the Hamiltonian is of the form $\hat\mathcal H_o=\hat\mathcal T+\hat\mathcal V$,
so the relationship between the superfluid density and the winding number
applies. A condition for the discretization to be valid is that the discretized Hamiltonian should reproduce
quantitatively the same physics as its continuous space analog when the lattice constants go to zero.
By performing the change of variables (\ref{Contravariant}), the continuous-space kinetic operator (\ref{KineticEnergy}) can be
rewritten as
\begin{equation}
  \hat\mathcal T=-\frac{\hbar^2|J|A^{-1}_{ik}A^{-1}_{jk}}{2m}\int_{\Omega}\!\!\hat\psi^\dagger(\vec q)\frac{\partial^2}{\partial q_i\partial q_j}\hat\psi(\vec q)\textrm{d}Q,
\end{equation}
where $|J|$ is the Jacobian determinant,
\begin{equation}
  |J|=\bigg|\frac{\partial(r^1,\cdots,r^d)}{\partial(q^1,\cdots,q^d)}\bigg|,
\end{equation}
and $\textrm{d}Q=\textrm{d}q_1\cdots\textrm{d}q_d$. Performing a second change of variables for each coordinates, $q_j=l_j y_j$, where $l_j$ has
the dimension of a length and $y_j$ is the new dimensionless variable, the kinetic operator becomes
\begin{equation}
  \hat\mathcal T=-\frac{\hbar^2|J|A^{-1}_{ik}A^{-1}_{jk}}{2ml_i l_j}\int_{\Omega}\!\! a^\dagger_{\vec y}\frac{\partial^2}{\partial y_i\partial y_j}a_{\vec y}^{\phantom\dagger}\:\textrm{d}y_1\cdots\textrm{d}y_d,
\end{equation}
where we have used the previously defined dimensionless creation and annihilation operators, (\ref{DiscreteField1}) and (\ref{DiscreteField2}).
As before, in the limit $l_j\to 0$, the integral becomes independent of the step size $\textrm{d}y_j$, which can be chosen
as unity. In this case, the integral becomes discrete, $l_j$ becomes
the lattice constant in the $\vec q_j$ direction, and the kinetic operator takes the form:
\begin{equation}
  \hat\mathcal T=-\frac{\hbar^2|J|A^{-1}_{ik}A^{-1}_{jk}}{2ml_i l_j}\sum_{\vec y\in\Omega}a^\dagger_{\vec y}\frac{\partial^2}{\partial y_i\partial y_j}a_{\vec y}^{\phantom\dagger}
\end{equation}
It is important to keep in mind that the volume $\Omega$ is a hyperrectangle. However, if the basis $\mathcal B_q$ is not orthogonal,
summing over a hyperparallelepiped turns out to be more convenient. As a result, instead of a hyperrectangle of volume
$\Omega=\prod_j L_j$, we consider a hyperparallelepiped of volume $\tilde\Omega=|J|\prod_j L_j$. Since the volume $\tilde\Omega$
is scaled by a factor $|J|$ with respect to $\Omega$, the same energy can be recovered by multiplying it by the inverse factor.
Therefore, the Jacobian determinant disappears and the kinetic operator is equivalent to
\begin{equation}
  \label{GeneralDiscreteKinetic} \hat\mathcal T=-\frac{\hbar^2A^{-1}_{ik}A^{-1}_{jk}}{2ml_i l_j}\sum_{\vec y\in\tilde\Omega}a^\dagger_{\vec y}\frac{\partial^2}{\partial y_i\partial y_j}a_{\vec y}^{\phantom\dagger},
\end{equation}
where the summation is over all vectors $\vec y$ in the volume $\tilde\Omega$ with the components $y_j$ varying over $S_j=L_j/l_j$ lattice sites.
We can use for the second-order derivative the previous symmetrical prescription (\ref{SecondPrescription}).
The discretization of the potential term $\hat\mathcal V$ can be done in a similar way, and does not affect the
following conclusions.

\subsection{Application to Bravais lattices}
We give here some examples of application of the above discretization to some common Bravais lattices, and we emphasize
the differences between our expressions for the superfluid density and the expressions that are usually improperly used.
For simplicity, we consider here only isotropic cases.
\subsubsection{Hypercubic lattice}
In the case of a hypercubic lattice with $S_1\times\cdots\times S_d$ sites, the basis of the primitive cell is orthogonal, and
the lattice constants $l_j$ are all equal to the same value~$l$. Using the identity transformation for $\mathcal A$ and
defining $t=\frac{\hbar^2}{2ml^2}$, Eq.~(\ref{GeneralDiscreteKinetic}) leads to the previous discrete form of the kinetic energy:
\begin{equation}
  \label{HypercubicKinetic}\hat\mathcal T=-t\sum_{\langle p,q\rangle}\big(a_p^\dagger a_q^{\phantom\dagger}+H.c.\big)+2td\hat\mathcal N
\end{equation}
In this simple case, the metric tensor $g_{\mu\nu}$ is just the identity, and combining (\ref{PollockCeperleySuperfluid})
and (\ref{DimensionlessRhoS}) leads to
\begin{equation}
  \label{HypercubicRhoS} \tilde\rho_s=\frac{S^{2-d}}{2t\beta d}\langle\mathcal W_1^2+\cdots+\mathcal W_d^2\rangle,
\end{equation}
where we have assumed the same number of lattice sites $S_j=S$ in each of the primitive directions. For this case, we recover the
well-known expression. A common mistake arises when applying (\ref{HypercubicRhoS}) to non-cubic lattice geometries, as we show below.

\subsubsection{Triangular lattice}
\begin{figure}[h]
  \centerline{\includegraphics[width=0.35\textwidth]{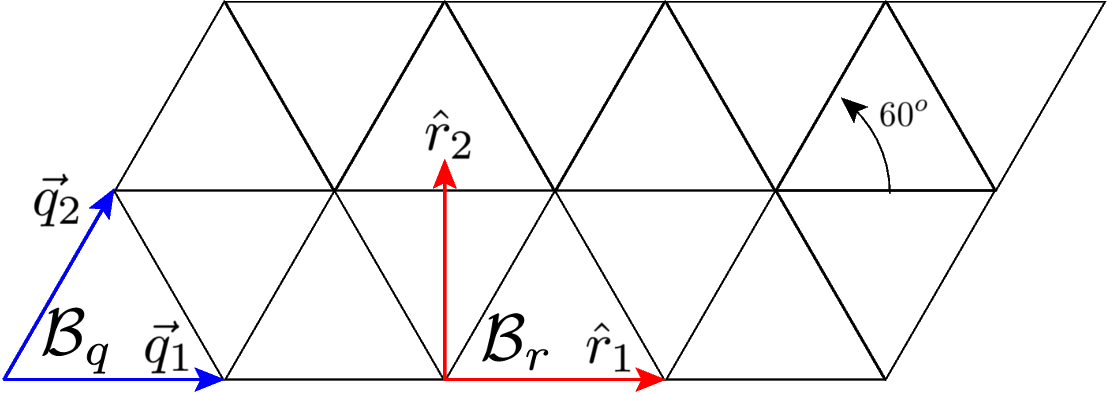}}
  \caption
    {
      (Color online) Triangular lattice. The basis $\mathcal B_r$ is changed to $\mathcal B_q$ by the transformation (\ref{TriangularTransform}).
    }
  \label{TriangularLattice}
\end{figure}

We address here the case of a $S\times S$ triangular lattice (Fig.~\ref{TriangularLattice}). The transformation matrix $A$ that
changes the orthonormal basis $\mathcal B_r=\lbrace\hat r_1,\hat r_2\rbrace$ into the basis $\mathcal B_q=\lbrace\vec q_1,\vec q_2\rbrace$
and the metric tensor $g_{\mu,\nu}$ are given by:
\begin{equation}
  \label {TriangularTransform} A=\left(\begin{array}{cc}1&1/2\\
                           0&\sqrt{3}/2\end{array}\right),\quad
  g=\left(\begin{array}{cc}1&1/2\\
                          1/2&1\end{array}\right)
\end{equation}
\newline
With this transformation, Eq.~(\ref{GeneralDiscreteKinetic}) becomes
\begin{equation}
  \label{TriangularKinetic} \hat\mathcal T=-t^\prime\sum_{\langle p,q\rangle}\big(a_p^\dagger a_q^{\phantom\dagger}+H.c.\big)+6t^\prime\hat\mathcal N,
\end{equation}
with the energy scale $t^\prime=\frac{\hbar^2}{3ml^2}$. As a result, using (\ref{DotProduct}) for the square of the winding
number operator, the dimensionless superfluid density is:
\begin{equation}
  \label{TriangularRhoS} \tilde\rho_s=\frac{1}{6t^\prime\beta}\langle\mathcal W_1^2+\mathcal W_2^2+\mathcal W_1\mathcal W_2\rangle
\end{equation}
The above expression computed with the energy scale $t^\prime=1$ differs significantly from the quantity
(\ref{HypercubicRhoS}) that is usually improperly applied with $t=1$. Doing so not only introduces an
energy scale mismatch between the simulated Hamiltonian and the computed superfluid density, but some winding correlations are
missed too.

\subsubsection{Face-centered cubic lattice}
\begin{figure}[h]
  \centerline{\includegraphics[width=0.4\textwidth]{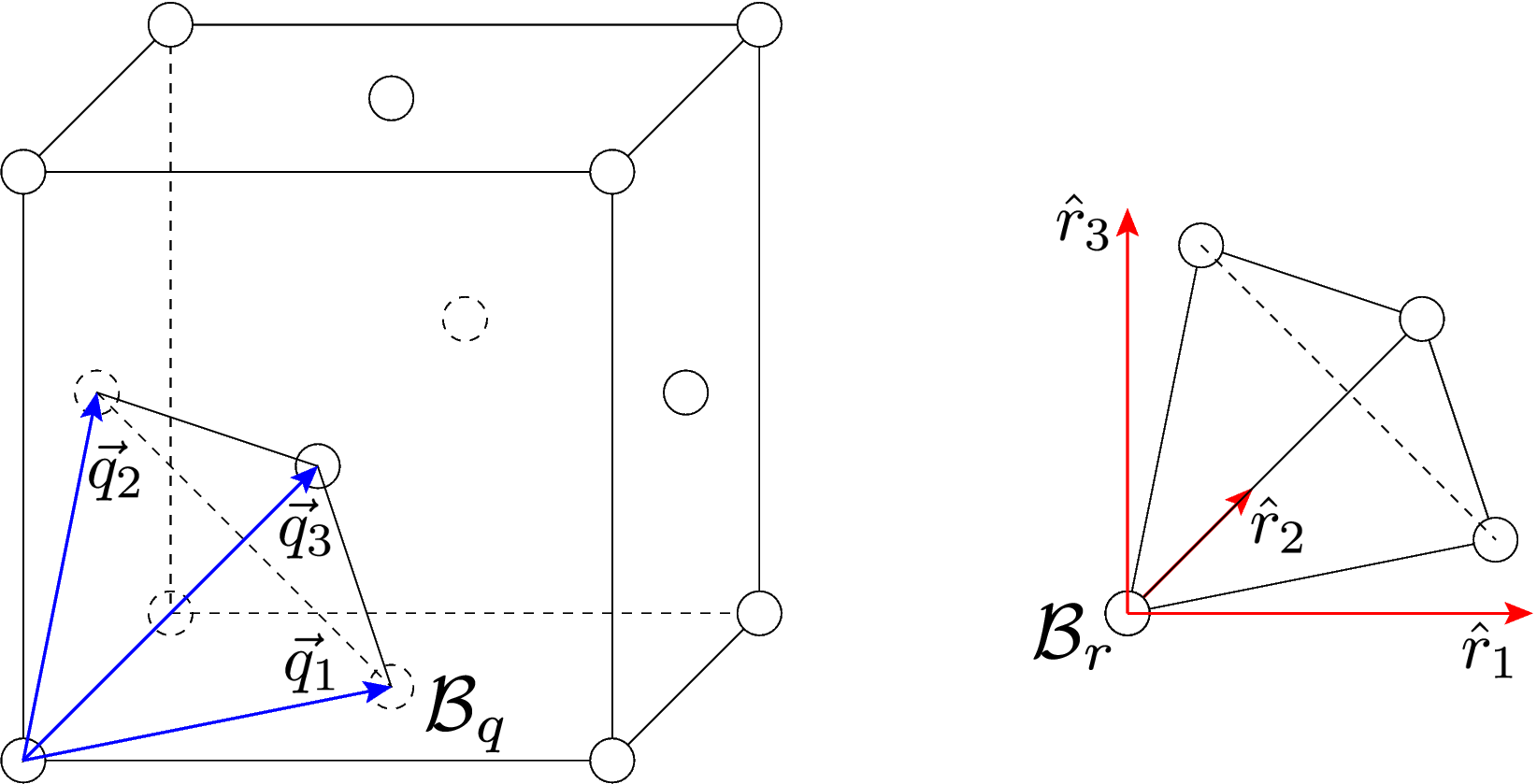}}
  \caption
    {
      (Color online) Face-centered cubic lattice. The basis $\mathcal B_r$ is changed to $\mathcal B_q$ by the transformation (\ref{FccTransform}).
    }
  \label{FccLattice}
\end{figure}

The transformation matrix $A$ and the metric tensor~$g_{\mu,\nu}$ associated to the primitive cell of a face-centered cubic lattice (Fig.~\ref{FccLattice})
are given by:
\begin{equation}
  \label{FccTransform} A=\frac{1}{\sqrt{2}}\left(\begin{array}{ccc}1&0&1\\
                                              1&1&0\\
                                              0&1&1\end{array}\right),\quad
  g=\left(\begin{array}{ccc}1&1/2&1/2\\
                            1/2&1&1/2\\
                            1/2&1/2&1\end{array}\right)
\end{equation}
\newline
With this transformation, Eq.~(\ref{GeneralDiscreteKinetic}) becomes
\begin{equation}
  \label{FccKinetic} \hat\mathcal T=-t^{\prime\prime}\sum_{\langle p,q\rangle}\big(a_p^\dagger a_q^{\phantom\dagger}+H.c.\big)+12t^{\prime\prime}\hat\mathcal N,
\end{equation}
with the energy scale $t^{\prime\prime}=\frac{\hbar^2}{4ml^2}$. Thus, for a $S\times S\times S$ lattice, the dimensionless superfluid
density takes the form:
\begin{equation}
  \label{FccRhoS} \tilde\rho_s=\frac{\langle\mathcal W_1^2+\mathcal W_2^2+\mathcal W_3^2+\mathcal W_1\mathcal W_2+\mathcal W_2\mathcal W_3+\mathcal W_3\mathcal W_1\rangle}{12t^{\prime\prime}\beta S}
\end{equation}
Once again, the above expression computed with the energy scale $t^{\prime\prime}=1$ differs significantly from the quantity~(\ref{HypercubicRhoS})
that is usually improperly applied with $t=1$.

\subsection{Application to non-Bravais lattices}
A non-Bravais lattice can be described as a basis of points that is reproduced at each point of an underlying Bravais lattice.
Another possible description is to consider it as a Bravais lattice with smaller lattice constants and missing points.
The advantage of this latter description is that we already know how to discretize continuous space and obtain a Bravais lattice
with the associated expression of the superfluid density. By adding to the Hamiltonian an infinite potential at the locations of
the missing points, we can prevent the particles from occupying those positions and generate the corresponding non-Bravais lattice.
This mathematical ``trick" allows us to determine the expression of the superfluid density.

\subsubsection{Honeycomb lattice}
\begin{figure}[h]
  \centerline{\includegraphics[width=0.45\textwidth]{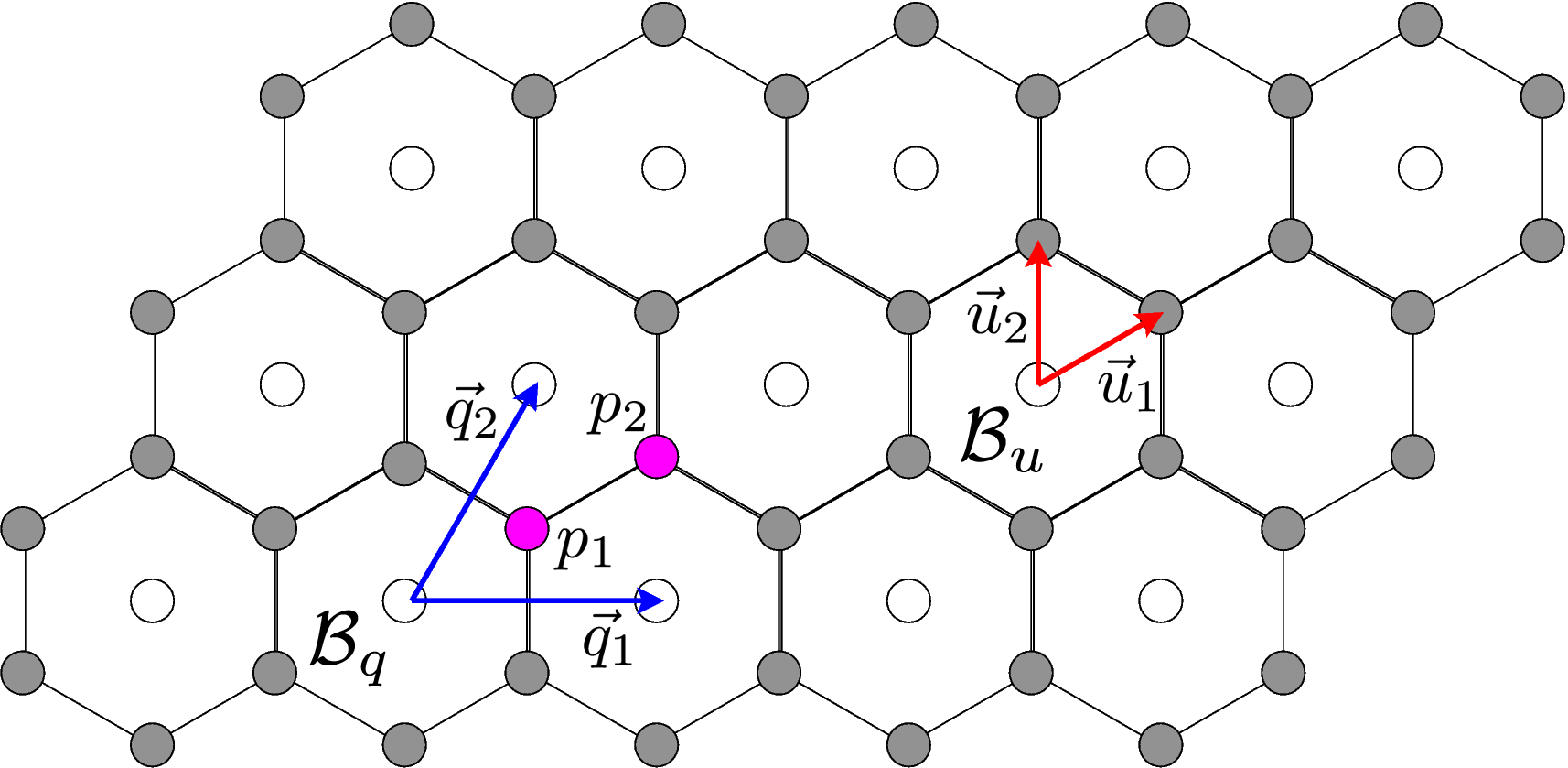}}
  \caption
    {
      (Color online) A honeycomb lattice can be described as a two-point basis $(p_1,p_2)$ that is reproduced at each point of
      a triangular lattice generated by a basis $\mathcal B_q$, or as a triangular lattice generated by a basis $\mathcal B_u$
      to which all points generated by the basis $\mathcal B_q$ are removed.
    }
  \label{HoneycombLattice}
\end{figure}

A honeycomb lattice (Fig.~\ref{HoneycombLattice}) is usually seen as a two-point basis $(p_1,p_2)$ that is reproduced at each point of
a triangular lattice generated by a basis $\mathcal B_q$. In our case, it is more convenient to describe it as a triangular
lattice generated by a second basis $\mathcal B_u$, to which we remove all points generated by the basis $\mathcal B_q$.

At this point, it is useful to consider the Hamiltonian $\hat\mathcal H=\hat\mathcal T+\hat\mathcal V$ with $\hat\mathcal T$
given by (\ref{KineticEnergy}) and $\hat\mathcal V$ by
\begin{equation}
  \label{PotentialOperator}\hat\mathcal V=\int_\Omega\hat\psi^\dagger(\vec r)V(\vec r)\hat\psi(\vec r)\textrm{d}\Omega,
\end{equation}
with
\begin{equation}
  \label{PotentialFunction}V(\vec r)=Uh^2\sum_{j_1,j_2}\delta(\vec r-j_1\vec q_1-j_2\vec q_2),
\end{equation}
where $U$ is a parameter with the dimension of an energy and $h=||\vec q_1||=||\vec q_2||$. Injecting (\ref{PotentialFunction}) into
(\ref{PotentialOperator}) and using the previously defined dimensionless creation and annihilation operators (\ref{DiscreteField1})
and (\ref{DiscreteField2}), the potential becomes:
\begin{equation}
  \hat\mathcal V=U\sum_{j_1,j_2}\hat n_{j_1\vec q_1+j_2\vec q_2}
\end{equation}
By discretizing $\hat\mathcal T$ with the transformation that changes an orthogonal basis into the basis $\mathcal B_u$ and
defining $l=||\vec u_1||=||\vec u_2||$, the Hamiltonian becomes
\begin{eqnarray}
  \nonumber\hat\mathcal H &=& -t'\sum_{\langle p,q\rangle}(a_p^\dagger a_q^{\phantom\dagger}+H.c.\big)+6t'\hat\mathcal N\\
                          && +U\sum_{p\in\mathcal B_q}\hat n_p
\end{eqnarray}
where the sum $\sum_{\langle p,q\rangle}$ is over all distinct pairs of first neighboring sites of the triangular lattice generated
by~$\mathcal B_u$, and the sum $\sum_{p\in\mathcal B_q}$ is over all sites generated by $\mathcal B_q$. Since the Hamiltonian $\hat\mathcal H$ satisfies the condition (\ref{ParticularClass}),
the corresponding superfluid density is given by (\ref{TriangularRhoS}), and this result applies for any value of the parameter $U$.
In particular, it applies in the limit $U\to\infty$ where the Hamiltonian becomes equivalent to
\begin{equation}
  \hat\mathcal H=-t'\sum_{\langle p,q\rangle-{\mathcal B_q}}(a_p^\dagger a_q^{\phantom\dagger}+H.c.\big)+6t'\hat\mathcal N,
\end{equation}
where the notation $\langle p,q\rangle-{\mathcal B_q}$ indicates that the points generated by $\mathcal B_q$
are removed. As a result, the above Hamiltonian describes particles on a honeycomb lattice, and the expression of the superfluid density
is the same as for a triangular lattice and given by (\ref{TriangularRhoS}).

\subsubsection{Kagome lattice}
\begin{figure}[h]
  \centerline{\includegraphics[width=0.45\textwidth]{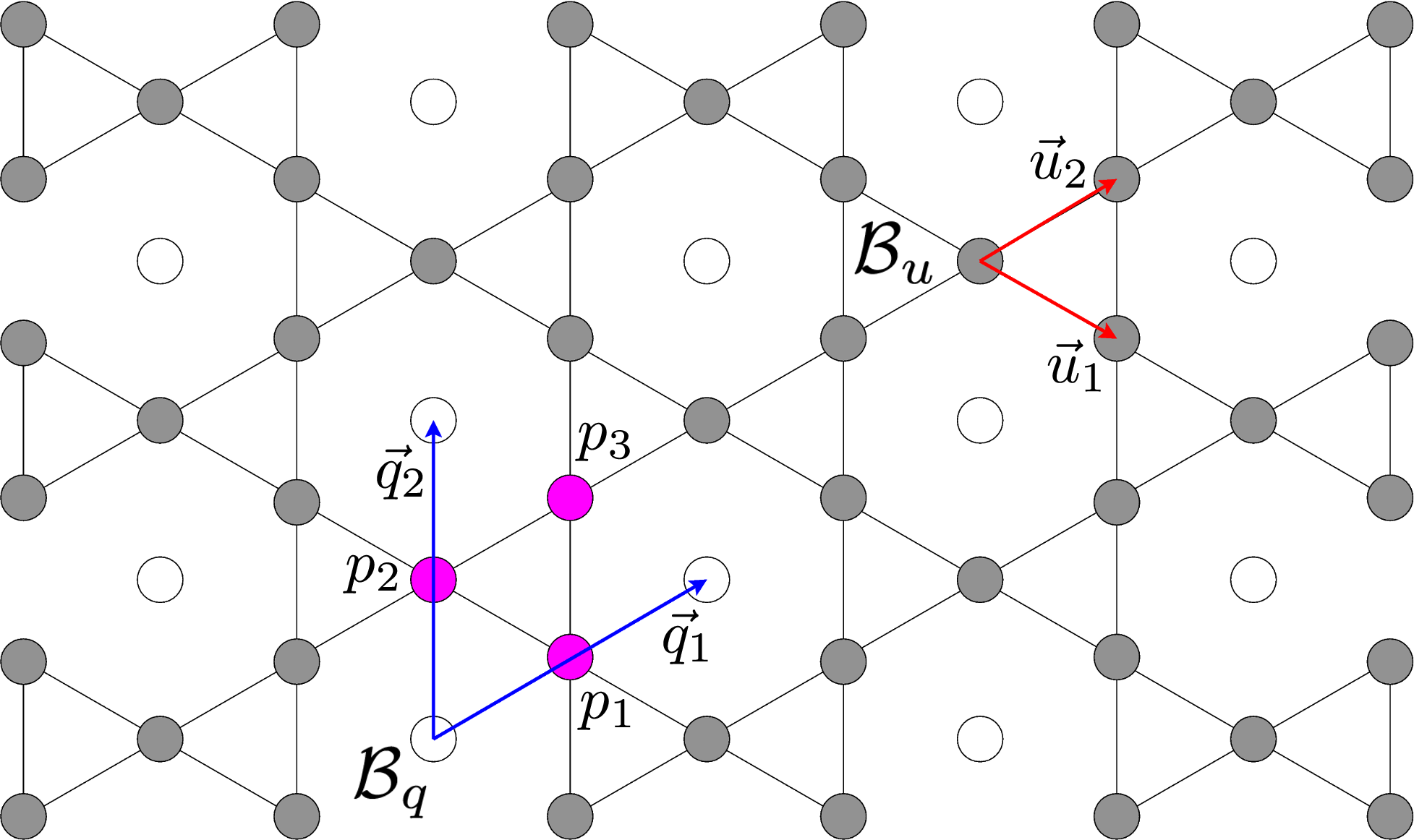}}
  \caption
    {
      (Color online) A kagome lattice can be described as a three-point basis $(p_1,p_2,p_3)$ that is reproduced at each point of
      a triangular lattice generated by a basis $\mathcal B_q$, or as a triangular lattice generated by a basis $\mathcal B_u$
      to which all points generated by the basis $\mathcal B_q$ are removed.
    }
  \label{KagomeLattice}
\end{figure}

A kagome lattice (Fig.~\ref{KagomeLattice}) is formed by corner-sharing triangles, and is usually seen as a three-point basis $(p_1,p_2,p_3)$ that is reproduced at each point of
a triangular lattice generated by a basis $\mathcal B_q$. Here again, it is more convenient to describe it as a triangular
lattice generated by a second basis $\mathcal B_u$, to which we remove all points generated by the basis $\mathcal B_q$.

Therefore, the same reasoning as for the honeycomb lattice can be applied, and we conclude that the expression of the superfluid
density is given by that of a triangular lattice (\ref{TriangularRhoS}).

\subsubsection{Pyrochlore lattice}
\begin{figure}[h]
  \centerline{\includegraphics[width=0.45\textwidth]{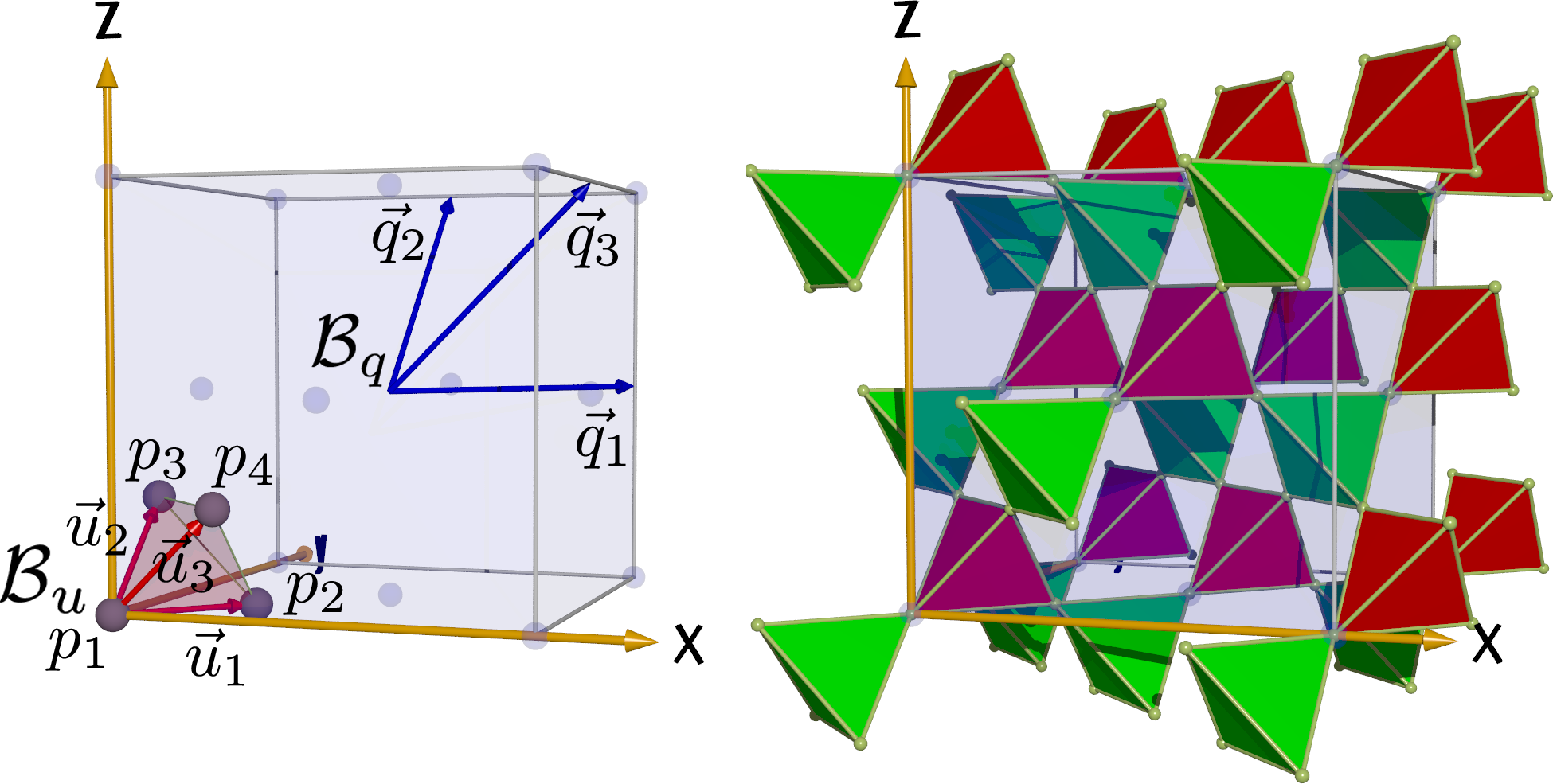}}
  \caption
    {
      (Color online) A pyrochlore lattice can be described as a four-point basis $(p_1,p_2,p_3,p_4)$ that is reproduced at each point of
      a face-centered cubic lattice generated by a basis $\mathcal B_q$, or as a face-centered cubic lattice generated by a basis $\mathcal B_u$
      to which the points generated by the basis $\mathcal B_q$ are removed.
    }
  \label{PyrochloreLattice}
\end{figure}

A pyrochlore lattice (Fig.~\ref{PyrochloreLattice}) is formed by corner-sharing tetrahedrons, and is usually seen as a four-point basis $(p_1,p_2,p_3,p_4)$ that is reproduced at each point of
a face-centered cubic lattice generated by a basis $\mathcal B_q$. In a way similar to the honeycomb and kagome lattices, it is more convenient to describe it as a
face-centered cubic lattice generated by a second basis $\mathcal B_u$, to which we remove all points generated by the basis $\mathcal B_q$.

As before, the same reasoning as for the honeycomb and kagome lattices can be applied, and we conclude that the expression of the superfluid
density is given by that of a face-centered cubic lattice (\ref{FccRhoS}).

\subsection{Consistency check}
In this subsection, we make a consistency check that illustrates the correctness of our expressions of the superfluid density
for the hypercubic (\ref{HypercubicRhoS}), triangular (\ref{TriangularRhoS}), and face-centered cubic ({\ref{FccRhoS}) lattices.
Since the kinetic term of the triangular lattice (\ref{TriangularKinetic}) and the kinetic term of the face-centered cubic (\ref{FccKinetic})
lattice correspond to the discretization of the continuous kinetic term (\ref{KineticEnergy}) with $d=2$ and $d=3$, respectively, they should give exactly the same
superfluid density as the kinetic term of the hypercubic lattice (\ref{HypercubicKinetic}) with the corresponding dimensionality.

In order to check this, we made use of the Stochastic Green Function\cite{SGF} (SGF) algorithm with directed updates\cite{DirectedSGF}, and performed quantum
Monte Carlo simulations of the kinetic term for hard-core bosons at half-filling. The results are shown in Fig.~\ref{ConsistencyCheck}.
The superfluid density obtained for a $16\times 16$ triangular lattice with $t'=1$ is in agreement with the superfluid density obtained
for a $16\times 16$ square lattice with $t=1$, as a function of temperature $T$, the small differences being due to finite-size effects. We get the same agreement between the superfluid
density obtained for a $4\times4\times4$ face-centered cubic lattice with $t''=1$ and the superfluid density obtained for a
$4\times4\times4$ cubic lattice with $t=1$.
\begin{figure}[h]
  \centerline{\includegraphics[width=0.45\textwidth]{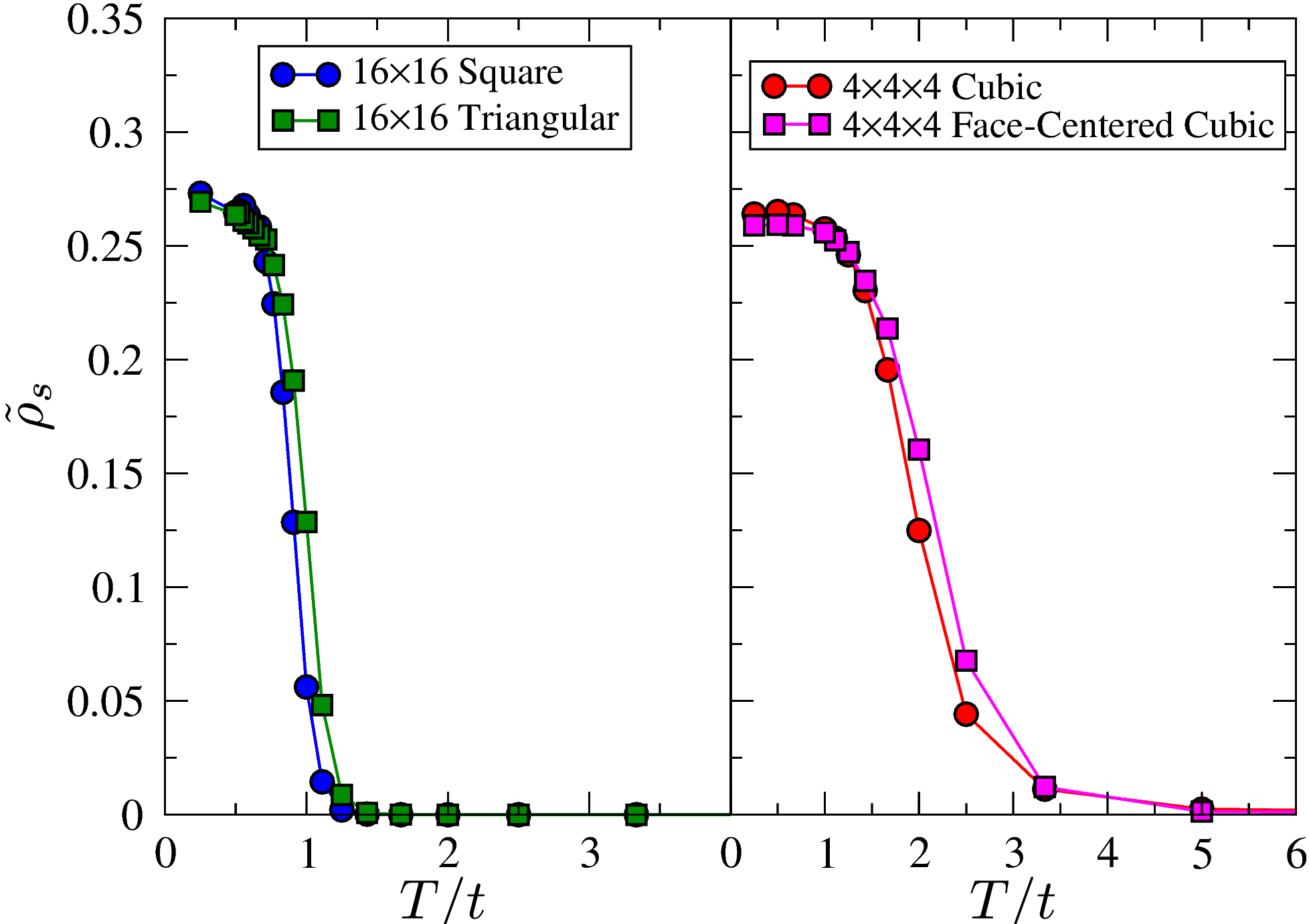}}
  \caption
    {
      (Color online) The dimensionless superfluid density $\tilde\rho_s$ of hard-core bosons as a function of temperature $T/t$ at half-filling, for different lattice
      geometries. The values obtained for a triangular lattice agree with the values obtained for a square lattice, the small differences
      being due to finite-size effects. In the same way, the values obtained for a face-centered cubic lattice agree with the
      values obtained for a cubic lattice. The errorbars are smaller than the symbols' size.
    }
  \label{ConsistencyCheck}
\end{figure}

\section{\label{Example} Lattice Hamiltonian with hopping between second neighbors}
In this section, we illustrate the usefulness of Eq.~(\ref{DimensionlessGeneralRhoS}) by considering a Hamiltonian for which the well-known
expressions of the superfluid density are not applicable. The model consists of soft-core bosons on a two-dimensional $S\times S$ square lattice
described by the Hamiltonian:
\begin{eqnarray}
  \nonumber        \hat\mathcal H_o &=& -t\sum_{\langle p,q\rangle}\big(a_p^\dagger a_q^{\phantom\dagger}+H.c.\big)+\frac{U}{2}\sum_{j}\hat n_j(\hat n_j-1)\\
  \label{SecondNeighborHamiltonian} && -\lambda\sum_{\langle\langle p,q\rangle\rangle}\big(a_p^\dagger a_q^{\phantom\dagger}+H.c.\big)
\end{eqnarray}
The sum $\sum_{\langle\langle p,q\rangle\rangle}$ is over all distinct pairs of second-neighboring sites $p$ and $q$.
In addition to the discrete operators (\ref{DiscreteN}), (\ref{DiscreteR}), (\ref{DiscreteP}), (\ref{DiscreteT}),
we define the operator:
\begin{equation}
  \vec{\mathcal Q}=-i\sqrt{mt/2}\sum_{\langle\langle p,q\rangle\rangle}\big(a_p^\dagger a_q^{\phantom\dagger}-H.c.\big)(\vec q-\vec p)
\end{equation}
Calculating the commutator $\big[\vec{\mathcal R},\hat\mathcal H_o\big]$ leads to:
\begin{equation}
  \label{Commutator}\big[\vec{\mathcal R},\hat\mathcal H_o\big]=i\frac{\hbar}{m}\Big(\vec{\mathcal P}+\frac{\lambda}{t}\vec{\mathcal Q}\Big)
\end{equation}
Therefore, the Hamiltonian (\ref{SecondNeighborHamiltonian}) does not belong to the class defined by (\ref{ParticularClass}),
and the expression of the superfluid density given by (\ref{HypercubicRhoS}) with $d=2$ is not applicable. Injecting (\ref{Commutator})
into (\ref{DimensionlessGeneralRhoS}) with $S_{\textrm{\tiny tot}}=S^2$, we get the expression:
\begin{eqnarray}
  \nonumber  \tilde\rho_s     &=& \tilde\rho-\frac{1}{4tS^2}\big\langle\hat\mathcal T\big\rangle\\
  \label{SecondNeighborRhoS2} && -\frac{1}{2mS^2}\Big\langle\vec{\mathcal P}\cdot\!\!\int_0^\beta\!\! e^{\tau\hat\mathcal H_o}\Big(\vec{\mathcal P}+\frac{\lambda}{t}\vec{\mathcal Q}\Big)e^{-\tau\hat\mathcal H_o}\textrm{d}\tau\Big\rangle
\end{eqnarray}
Using the SGF algorithm\cite{SGF} with directed updates\cite{DirectedSGF}, it is easy to evaluate (\ref{SecondNeighborRhoS2})
for a given configuration of the particle worldlines by defining $n_\zeta$ as the number of hoppings in the direction $\zeta$,
with $\zeta=\leftarrow,\rightarrow,\uparrow,\downarrow,\swarrow,\nearrow,\searrow,\nwarrow$, the notation being self-explanatory. Then (\ref{SecondNeighborRhoS2}) takes the form:
\begin{eqnarray}
  \nonumber\tilde\rho_s       &=& \frac{1}{4t\beta S^2}\big\langle\!(n_{_\leftarrow}\!-\!n_{_\rightarrow})\!(n_{_\leftarrow}\!-\!n_{_\rightarrow}\!+\!n_{_\swarrow}\!-\!n_{_\nearrow}\!-\!n_{_\searrow}\!+\!n_{_\nwarrow})\\
  \label{SecondNeighborRhoS3} && +(n_{_\downarrow}-n_{_\uparrow})(n_{_\downarrow}-n_{_\uparrow}+n_{_\swarrow}-n_{_\nearrow}+n_{_\searrow}-n_{_\nwarrow})\big\rangle
\end{eqnarray}
We have simulated the Hamiltonian
(\ref{SecondNeighborHamiltonian}) with $t=1$, $\lambda=0.8$, and $U=20$, at half-filling ($\tilde\rho=\frac12$) as a function of
temperature. Fig.~\ref{SecondHopping} shows a comparison between the quantity given by the discrete form of Pollock and
Ceperley's formula (\ref{HypercubicRhoS}) and our expression (\ref{SecondNeighborRhoS3}). This example clearly demonstrates
that (\ref{HypercubicRhoS}) is not applicable for this Hamiltonian, since it gives a value that is greater than the total density.
On the other hand, our expression (\ref{SecondNeighborRhoS3}) ensures that $\tilde\rho_s\in[0;\tilde\rho]$.
\begin{figure}[h]
  \centerline{\includegraphics[width=0.45\textwidth]{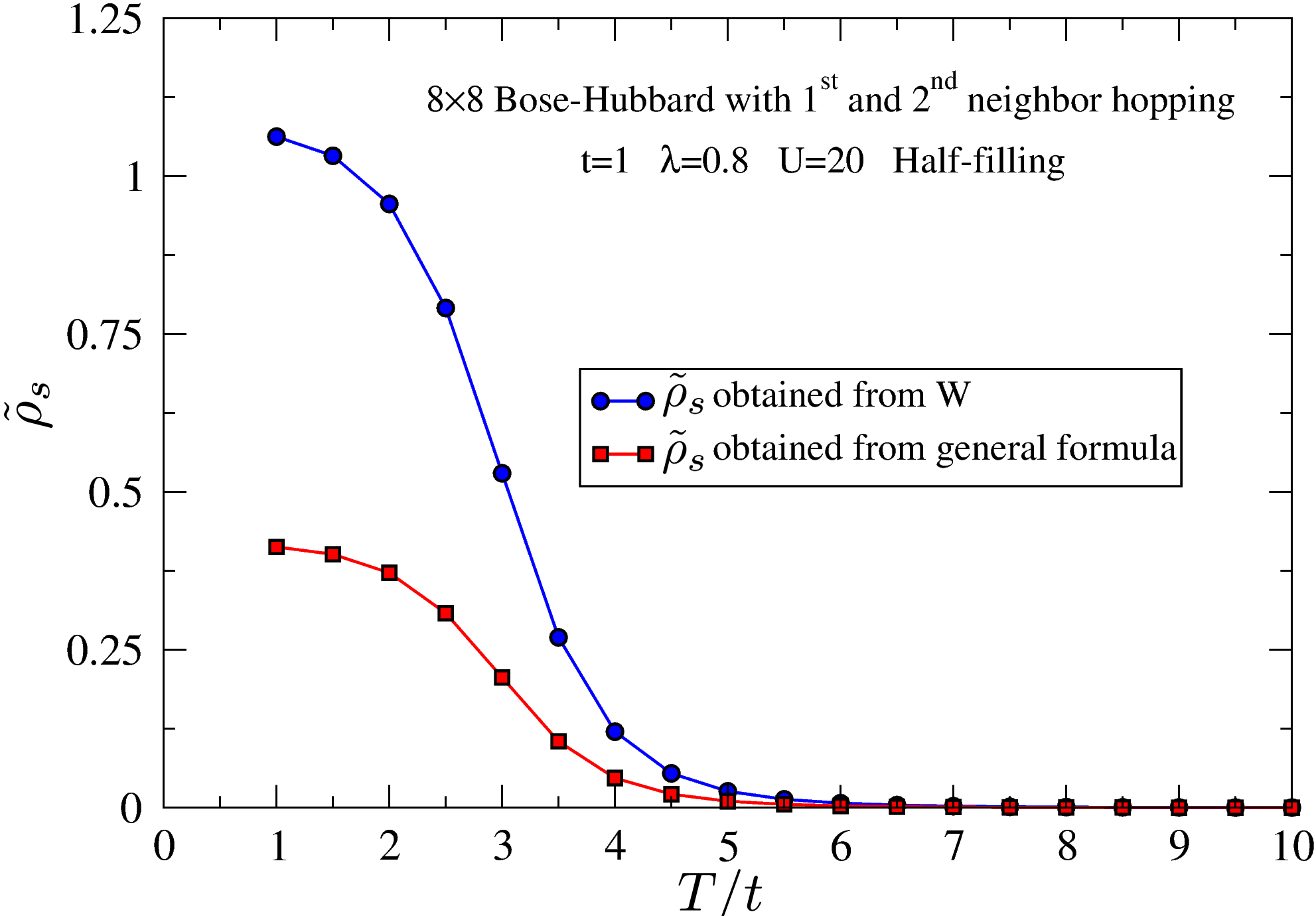}}
  \caption
    {
      (Color online) Comparison between the superfluid density (red) obtained from the general expression (\ref{SecondNeighborRhoS3})
      and the quantity (blue) given by (\ref{HypercubicRhoS}) for the Hamiltonian (\ref{SecondNeighborHamiltonian}). The errorbars
      are smaller than the symbols' size.
    }
  \label{SecondHopping}
\end{figure}

\section{Multi-species Hamiltonians}
The theory developed in sections \ref{Section3} and \ref{Section4} can be extended to multi-species Hamiltonians in order to
obtain the superfluid density of each component of mixtures. Again, for the sake of simplicity, we consider in the
following only the isotropic case. Consider a $d$-dimensional Hamiltonian $\hat\mathcal H_o$ with several species of particles. We denote by $m_\zeta$
the mass of a particle of a given species $\zeta$.

\subsection{Continuous space}
By adding the index $\zeta$ to the field operators that appear in (\ref{NumberOperator}), (\ref{PositionOperator}), and (\ref{MomentumOperator}), we can define
the continuous-space number $\hat\mathcal N_\zeta$, position $\vec{\mathcal R}_\zeta$, and momentum $\vec{\mathcal P}_\zeta$ operators associated
to each species $\zeta$.
As shown by Andreev and Bashkin\cite{Andreev}, the superfluid current $\vec j_s^\zeta$ of a
given species $\zeta$ can be carried by the other species. As a result, the superfluid density is a second order tensor,
and the superfluid current is given by
\begin{equation}
  \vec j_s^\zeta=\rho_s^{\zeta\xi}\vec v_s^{\:\xi},
\end{equation}
where $\vec v_s^{\:\xi}$ is the superfluid velocity of species $\xi$.
The friction between the normal components of each species imposes the normal velocity to be the same for all species. However
the different species can have different superfluid velocities.
For each species $\zeta$, we denote by $\mathcal F^\zeta$ the frame in which its supefluid component comes to rest,
and we denote by $\mathcal F'$ the frame of the moving walls in which the normal components of all species are at rest.
Defining $\vec v^{\:\zeta}$ as the velocity of $\mathcal F'$ with respect to $\mathcal F^\zeta$, we can define the unitary operator
\begin{equation}
  \hat\mathcal U=e^{-\frac{i}{\hbar}\sum_\zeta m_\zeta\vec v^{\:\zeta}\cdot\vec{\mathcal R}_\zeta},
\end{equation}
and interpret it as the operator that performs for each species $\zeta$ a Galilean transformation from $\mathcal F^\zeta$ to $\mathcal F'$.
Applying the correspondence principle in the frame $\mathcal F'$ of the moving walls, the quantum average of the momentum
operator of species $\zeta$ must be equal to the classical momentum, which is due to the superfluid only. In this frame, the
superfluid velocity of species $\xi$ is $\vec v_s^{\:\xi}=-\vec v^{\:\xi}$. Thus we have:
\begin{eqnarray}
  \nonumber -\rho_s^{\zeta\xi}\vec v^{\:\xi}\Omega &=& \frac{1}{\mathcal Z_o}\textrm{Tr }\vec\mathcal P_\zeta e^{-\beta\hat\mathcal H_o}\\
  \nonumber                                        &=& \frac{1}{\mathcal Z_o}\textrm{Tr }\hat\mathcal U^\dagger\vec\mathcal P_\zeta\hat\mathcal U e^{-\beta\hat\mathcal U^\dagger\hat\mathcal H_o\hat\mathcal U}\\
  \label{CorrespondencePrinciple2}                 &=& \frac{1}{\mathcal Z_o}\textrm{Tr }\big(\vec\mathcal P_\zeta-m_\zeta\vec v^{\:\zeta}\hat\mathcal N_\zeta\big)e^{-\beta\hat\mathcal U^\dagger\hat\mathcal H_o\hat\mathcal U}
\end{eqnarray}
Calculating the divergence with respect to $\vec v^{\:\xi}$ and taking the limit where all velocities go to zero, we get the
expression of the superfluid tensor
\begin{equation}
  \rho_s^{\zeta\xi}=\rho^\zeta\delta_{\zeta\xi}+i\frac{m_\xi}{\hbar\Omega d}\Big\langle\vec{\mathcal P}_\zeta\cdot\!\!\int_0^\beta\!\!\!\! e^{\tau\hat\mathcal H_o}\big[\vec{\mathcal R}_\xi,\hat\mathcal H_o\big]e^{-\tau\hat\mathcal H_o}\textrm{d}\tau\Big\rangle,
\end{equation}
where $\rho^\zeta=m_\zeta\langle\hat\mathcal N_\zeta\rangle/\Omega$ is the total density of species $\zeta$.

\subsection{Discrete space}
As before, it is useful to write the kinetic energy of species $\zeta$ as a sum of contributions from the different directions,
$\hat\mathcal T_\zeta=\sum_j\hat\mathcal T_{j\zeta}$, and see how the momentum $\vec{\mathcal P}_\zeta$ transforms under $\hat\mathcal U$:
\begin{equation}
  \label{DiscreteUnitaryTransformation2}\hat\mathcal U^\dagger\vec{\mathcal P}_\zeta\hat\mathcal U=\vec{\mathcal P}_\zeta-m_\zeta\vec v^{\:\zeta}\hat\mathcal N_\zeta+\frac{m_\zeta}{2t_\zeta}\sum_j v_j^\zeta\hat j\hat\mathcal T_{j\zeta}+\vec{\mathcal O}(\vec v^2)
\end{equation}
Thus, new terms proportional to $\hat\mathcal T_{j\zeta}$ need to be subtracted from (\ref{CorrespondencePrinciple2}), leading to:
\begin{eqnarray}
  \nonumber \rho_s^{\zeta\xi}&=& \Big(\rho^\zeta-\frac{m_\zeta}{2t_\zeta\Omega d}\big\langle\hat\mathcal T_\zeta\big\rangle\Big)\delta_{\zeta\xi}\\
                             && +i\frac{m_\xi}{\hbar\Omega d}\Big\langle\vec{\mathcal P}_\zeta\cdot\!\!\int_0^\beta\!\!\!\! e^{\tau\hat\mathcal H_o}\big[\vec{\mathcal R}_\xi,\hat\mathcal H_o\big]e^{-\tau\hat\mathcal H_o}\textrm{d}\tau\Big\rangle
\end{eqnarray}
Our previous definition of the dimensionless superfluid density (\ref{DimensionlessRhoS}) can be generalized to the
superfluid density tensor as
\begin{equation}
  \label{DimensionlessSuperfluidTensor} \tilde\rho_s^{\zeta\xi}=\frac{\rho_s^{\zeta\xi}\Omega}{m_\xi S_{\textrm{\tiny tot}}},
\end{equation}
Defining the for each species $\zeta$ the associated dimensionless density $\tilde\rho^\zeta=\rho^\zeta\Omega/m_\zeta S_{\textrm{\tiny tot}}$,
we get:
\begin{eqnarray}
  \nonumber \tilde\rho_s^{\zeta\xi}      &=& \Big(\tilde\rho^\zeta-\frac{1}{2t_\zeta dS_{\textrm{\tiny tot}}}\big\langle\hat\mathcal T_\zeta\big\rangle\Big)\delta_{\zeta\xi}\\
  \label{DimensionlessSuperfluidTensor2} && +i\frac{1}{\hbar dS_{\textrm{\tiny tot}}}\Big\langle\vec{\mathcal P}_\zeta\cdot\!\!\int_0^\beta\!\!\!\! e^{\tau\hat\mathcal H_o}\big[\vec{\mathcal R}_\xi,\hat\mathcal H_o\big]e^{-\tau\hat\mathcal H_o}\textrm{d}\tau\Big\rangle\quad
\end{eqnarray}

\subsection{\label{Example2} Application to a two-species Hamiltonian with inter-species conversion terms}
We consider here a one-dimensional lattice Hamiltonian with $S$ sites that describes atoms and molecules with inter-species conversion terms\cite{Feshbach1,Feshbach2},
which takes the form
\begin{eqnarray}
  \nonumber\hat\mathcal H_o   &=& -t_a\sum_{\langle p,q\rangle}\big(a_p^\dagger a_q^{\phantom\dagger}+H.c.\big)-t_m\sum_{\langle p,q\rangle}\big(m_p^\dagger m_q^{\phantom\dagger}+H.c.\big)\\
  \nonumber                   && +\frac{U_{aa}}{2}\sum_p \hat n_p^a(\hat n_p^a-1)+\frac{U_{mm}}{2}\sum_p \hat n_p^m(\hat n_p^m-1)\\
  \nonumber                   && +U_{am}\sum_p\hat n_p^a\hat n_p^m+D\sum_p\hat n_p^m\\
  \label{FeshbachHamiltonian} && +\sigma\sum_p\big(a_p^\dagger a_p^\dagger m_p^{\phantom\dagger}+H.c.\big),
\end{eqnarray}
where $a_p^\dagger$ and $a_p^{\phantom\dagger}$ (resp. $m_p^\dagger$ and $m_p^{\phantom\dagger}$) are the creation and annihilation
operators of an atom (resp. a molecule) on site $p$. The operator $\hat n_p^a=a_p^\dagger a_p^{\phantom\dagger}$ (resp $\hat n_p^m=m_p^\dagger m_p^{\phantom\dagger}$)
counts the number of atoms (resp. molecules) on site $p$.
The last term in (\ref{FeshbachHamiltonian}) converts a molecule
into two atoms and vice-versa. As a result, this Hamiltonian does not conserve the number of atoms nor the number of molecules,
but we can define the total density as $\tilde\rho^{\textrm{\tiny tot}}=\tilde\rho^a+2\tilde\rho^m$ which is conserved.
In a path-integral representation, the non-conservation of the number of atoms and molecules means that the atomic and molecular worldlines can be broken (Fig.~\ref{Winding}). As pointed in ref.\cite{Eckholt},
this results in the impossibility to define winding numbers for atoms and for molecules that are topologically conserved.
Nevertheless, our general expression of the superfluid density tensor (\ref{DimensionlessSuperfluidTensor2}) does not rely on any definition of
the winding number, and can be easily calculated.
\begin{figure}[h]
  \centerline{\includegraphics[width=0.45\textwidth]{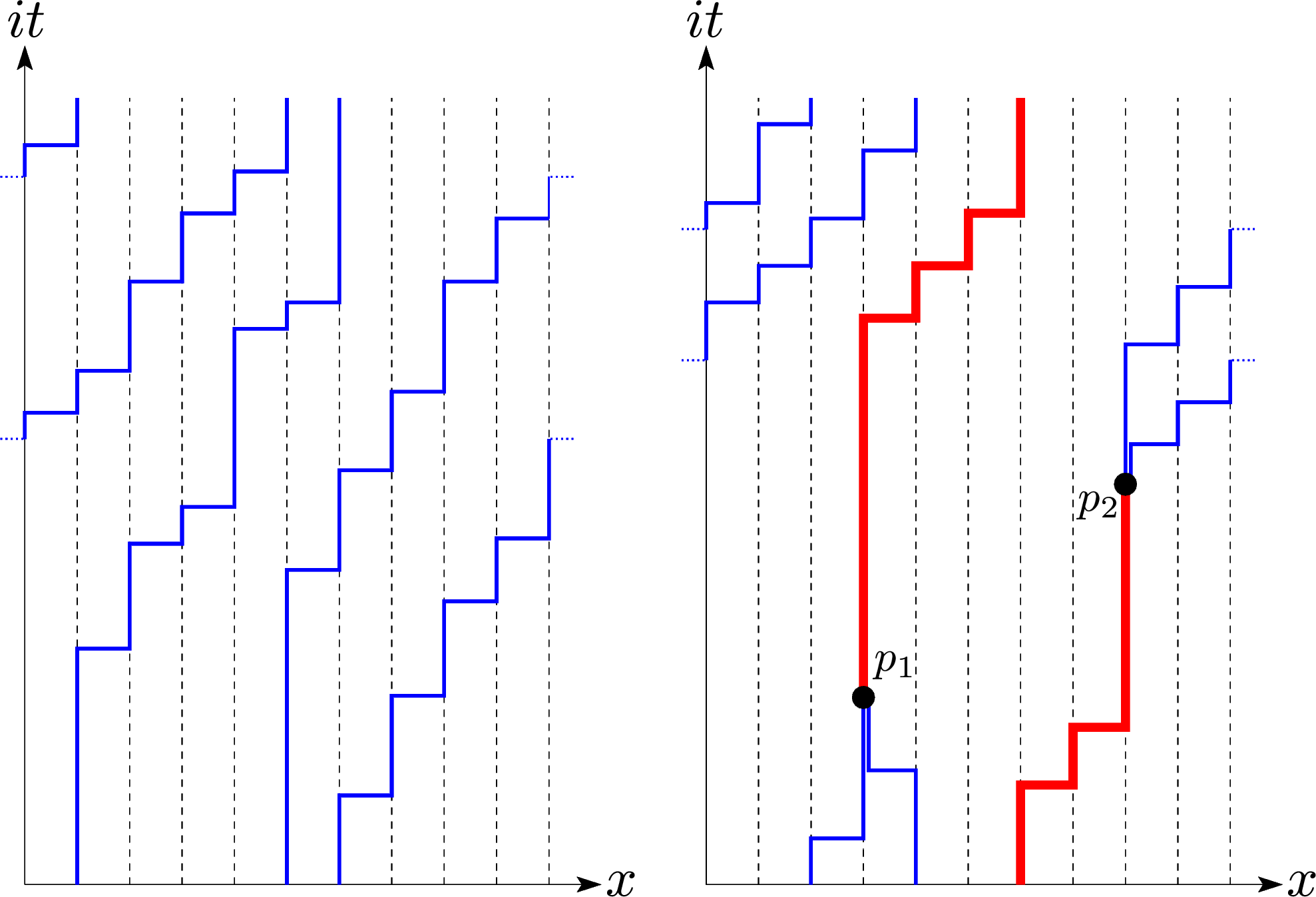}}
  \caption
    {
      (Color online) Worldline representation of a configuration of the partition function in the space and imaginary
      time $(x,it)$ plane. The left figure shows a configuration with 3 atoms (blue) when no molecules are formed. In this case,
      the winding number $W_a$ of the atoms can be defined, and the present realization corresponds to $W_a=2$. The right figure
      corresponds to a case when molecules (red) are formed. In this case, the numbers of atoms and molecules depend on imaginary
      time. Since the paths of the atoms and molecules are undefined between the points $p_1$ and $p_2$, it is not possible to assign
      winding numbers to them.
    }
  \label{Winding}
\end{figure}
Calculating the commutators of the position operators $\mathcal R_a$ and $\mathcal R_m$ with the Hamiltonian (\ref{FeshbachHamiltonian}), we obtain
\begin{eqnarray}
  && \label{Commutator5} \big[\mathcal R_a,\hat\mathcal H_o\big]=i\frac{\hbar}{m_a}\mathcal P_a+2\mathcal C,\\
  && \label{Commutator6} \big[\mathcal R_m,\hat\mathcal H_o\big]=i\frac{\hbar}{m_m}\mathcal P_m-\mathcal C,
\end{eqnarray}
where $\mathcal C$ is given by:
\begin{equation}
  \mathcal C=\sigma l\sum_p p\big(a_p^\dagger a_p^\dagger m_p^{\phantom\dagger}-H.c.\big)
\end{equation}
Injecting (\ref{Commutator5}) and (\ref{Commutator6}) in (\ref{DimensionlessSuperfluidTensor}), we obtain the elements of the superfluid density tensor:
\begin{eqnarray}
  \nonumber\tilde\rho_s^{aa} &=& \tilde\rho^a-\frac{1}{2t_a S}\big\langle\hat\mathcal T_a\big\rangle\\
  \nonumber                  && -\frac{1}{m_a S}\big\langle\mathcal P_a\int_0^\beta\mathcal P_a(\tau)\textrm{d}\tau\Big\rangle\\
  \label{RhoSaa}             && +\frac{2i}{\hbar S}\big\langle\mathcal P_a\int_0^\beta\mathcal C(\tau)\textrm{d}\tau\big\rangle
\end{eqnarray}
\begin{eqnarray}
  \nonumber\tilde\rho_s^{mm} &=& \tilde\rho^m-\frac{1}{2t_m S}\big\langle\hat\mathcal T_m\big\rangle\\
  \nonumber                  && -\frac{1}{m_m S}\big\langle\mathcal P_m\int_0^\beta\mathcal P_m(\tau)\textrm{d}\tau\Big\rangle\\
  \label{RhoSmm}             && -\frac{i}{\hbar S}\big\langle\mathcal P_m\int_0^\beta\mathcal C(\tau)\textrm{d}\tau\big\rangle
\end{eqnarray}
\begin{eqnarray}
  \nonumber\tilde\rho_s^{am} &=& -\frac{1}{m_m S}\big\langle\mathcal P_a\int_0^\beta\mathcal P_m(\tau)\textrm{d}\tau\Big\rangle\\
  \label{RhoSam}             && -\frac{i}{\hbar S}\big\langle\mathcal P_a\int_0^\beta\mathcal C(\tau)\textrm{d}\tau\big\rangle
\end{eqnarray}
\begin{eqnarray}
  \nonumber\tilde\rho_s^{ma} &=& -\frac{1}{m_a S}\big\langle\mathcal P_m\int_0^\beta\mathcal P_a(\tau)\textrm{d}\tau\Big\rangle\\
  \label{RhoSma}             && +\frac{2i}{\hbar S}\big\langle\mathcal P_m\int_0^\beta\mathcal C(\tau)\textrm{d}\tau\big\rangle
\end{eqnarray}
Evaluating (\ref{RhoSaa}), (\ref{RhoSmm}), (\ref{RhoSam}), and (\ref{RhoSma}) with the SGF method is made easy by defining $n_L^a$ and $n_R^a$ (resp. $n_L^m$ and $n_R^m$)
as the numbers of hoppings of atoms (resp. molecules) to the left and to the right in a given configuration of worldlines, and $n_{m\to a}^p$ and $n_{a\to m}^p$ as the
numbers of conversions of molecules to atoms and atoms to molecules that occur on site $p$. With these definitions, the elements of the superfluid
density tensor take the final forms:
\begin{eqnarray}
  \nonumber\tilde\rho_s^{aa} &=& \frac{1}{2t_a\beta S}\big\langle(n_L^a-n_R^a)^2\big\rangle\\
  \label{RhoSaa2}            && +\frac{1}{t_a\beta S}\big\langle(n_L^a-n_R^a)\sum_p p(n_{m\to a}^p-n_{a\to m}^p)\big\rangle\quad
\end{eqnarray}
\begin{eqnarray}
  \nonumber\tilde\rho_s^{mm} &=& \frac{1}{2t_m\beta S}\big\langle(n_L^m-n_R^m)^2\big\rangle\\
  \label{RhoSmm2}            && -\frac{1}{2t_m\beta S}\big\langle(n_L^m-n_R^m)\sum_p p(n_{m\to a}^p-n_{a\to m}^p)\big\rangle\quad
\end{eqnarray}
\begin{eqnarray}
  \nonumber\tilde\rho_s^{am} &=& \frac{1}{2t_a\beta S}\big\langle(n_L^a-n_R^a)(n_L^m-n_R^m)\big\rangle\\
  \label{RhoSam2}            && -\frac{1}{2t_a\beta S}\big\langle(n_L^m-n_R^m)\sum_p p(n_{m\to a}^p-n_{a\to m}^p)\big\rangle\quad
\end{eqnarray}
\begin{eqnarray}
  \nonumber\tilde\rho_s^{ma} &=& \frac{1}{2t_m\beta S}\big\langle(n_L^a-n_R^a)(n_L^m-n_R^m)\big\rangle\\
  \label{RhoSma2}            && +\frac{1}{t_m\beta S}\big\langle(n_L^a-n_R^a)\sum_p p(n_{m\to a}^p-n_{a\to m}^p)\big\rangle\quad
\end{eqnarray}
Figure \ref{Feshbach} shows the densities $\tilde\rho^a$ and $\tilde\rho^m$ of atoms and molecules, and the elements of
the superfluid density tensor obtained from (\ref{RhoSaa2}), (\ref{RhoSmm2}), (\ref{RhoSam2}), and (\ref{RhoSma2}) as functions
of the total density $\tilde\rho^{\textrm{\tiny tot}}$. With the parameters $S=20$, $t_a=1$, $t_m=0.5$, $\sigma=0.5$,
$U_{aa}=8$, $U_{mm}=100$, $U_{am}=12$, $D=6$, and $\beta=10$, our simulations indicate that the phase is incompressible for
densities $\tilde\rho^{\textrm{\tiny tot}}=1$ and $\tilde\rho^{\textrm{\tiny tot}}=2$, and nearly incompressible for $\tilde\rho^{\textrm{\tiny tot}}=3$.
This is consistent with the features observed in $\tilde\rho_s^{aa}$, $\tilde\rho_s^{mm}$, $\tilde\rho_s^{am}$, and
$\tilde\rho_s^{ma}$, and in agreement with ref.\cite{Eckholt}.
\begin{figure}[h]
  \centerline{\includegraphics[width=0.45\textwidth]{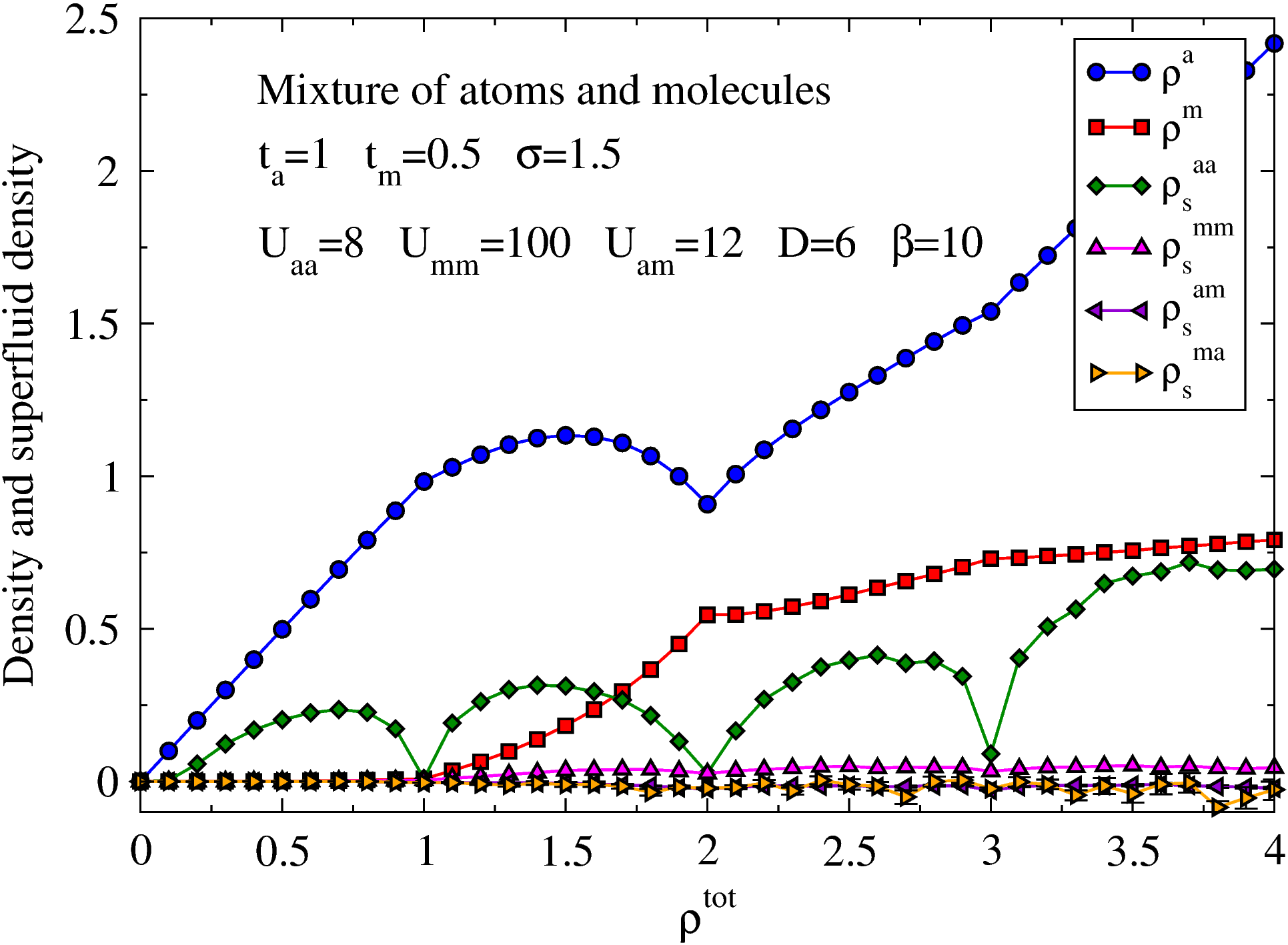}}
  \caption
    {
      (Color online) The densities $\tilde\rho^a$ and $\tilde\rho^m$ and the elements of the superfluid density tensor $\tilde\rho_s^{aa}$,
      $\tilde\rho_s^{mm}$,$\tilde\rho_s^{am}$, and $\tilde\rho_s^{ma}$, as functions of the total density $\tilde\rho^{\textrm{\tiny tot}}$. The errorbars are smaller than the symbols' size.
    }
  \label{Feshbach}
\end{figure}

\section{Conclusion}
Based on real and thought experiments, we give definitions of the superfluid density for general Hamiltonians, including
multi-species Hamiltonians. We derive general expressions that allow us to calculate the superfluid density
with path-integral methods. While it is well known that the superfluid density can be related to the response of the free
energy to a boundary phase twist or to the fluctuations of the winding number, we show that this is true only for a particular
class of Hamiltonians. Our expressions, however, can be applied to any Hamiltonian. In particular, they can be applied to Hamiltonians
that do not conserve the number of particles, where the winding number is undefined. By performing a
discretization of space with a general change of basis, we obtain formulae for the superfluid density for various lattice geometries. We point to some common
mistakes that occur when the energy scale is not correctly reflected in the expression of the superfluid density, and when
some correlations are missed because of a non-diagonal metric tensor. Finally, we give two examples of lattice Hamiltonians for which
the well-known expressions of the superfluid density are not applicable. We calculate the superfluid densities for these Hamiltonians
by evaluating our general expressions by means of quantum Monte Carlo simululations, using the SGF algorithm.

\begin{acknowledgments}
I would like to express special thanks to Mark Jarrell and Juana Moreno for providing support, and George Batrouni,
Fr\'ed\'eric H\'ebert, and Ka-Ming Tam for enlightening discussions. I am also grateful to Grisha Volovik for his useful
comments. This work is supported by NSF OISE-0952300.
\end{acknowledgments}


\begin{thebibliography}{1}
\bibitem{Kapitza} P.~Kapitza, Nature {\bf 141}, 74 (1938).
\bibitem{Allen} J.F.~Allen and A.D.~Misener, Nature {\bf 141}, 75 (1938).
\bibitem{Fisher} M.E.~Fisher, M.N.~Barber, and D.~Jasnow, Phys. Rev. A {\bf 8}, 2 (1973).
\bibitem{PollockCeperley} E.L.~Pollock and D.M.~Ceperley, Phys. Rev. B {\bf 36}, 8343 (1987).
\bibitem{Legget} A.J.~Leggett, in \textit{Topics in Superfluidity and Superconductivity, in Low Temperature Physics}, Proceedings of Blydepoort South Africa, 1991, edited by M.~J.~R.~Hoch and R.~H.~Lemmer (Springer–Verlag, 1991).
\bibitem{Scalapino} D.J.~Scalapino, S.R.~White, and S.C.~Zhang, PRL 68, 2830 (1992).
\bibitem{Batrouni} G.G.~Batrouni, Phys. Rev. B {\bf 70}, 184517 (2004).
\bibitem{Sorella} S.~Sorella, AIP Conference Proceedings, 2006, Vol. 816 Issue 1, p265.
\bibitem{Balazs1} Bal\'azs Het\'enyi, J. Phys. Soc. Jpn. 81, 124711 (2012).
\bibitem{Balazs2} Bal\'azs Het\'enyi, J. Phys. Soc. Jpn. 83, 034711 (2014).
\bibitem{Hess} G.B.~Hess and W.~M.~Fairbank, Phys. Rev. Lett. {\bf 19}, 216 (1967).
\bibitem{Whitmore} S.C.~Whitmore and W.~Zimmermann, Jr., Phys. Rev. Lett. {\bf 15}, 389 (1965).
\bibitem{Ekholm} D.T.~Ekholm and R.~B.~Hallock, Phys. Rev. B {\bf 21}, 3902 (1980).
\bibitem{Onsager} L. Onsager, Nuovo Cimento, Suppl. 6, 249 (1949).
\bibitem{SGF} V.G.~Rousseau, Phys. Rev. E {\bf 77}, 056705 (2008).
\bibitem{DirectedSGF} V.G.~Rousseau, Phys. Rev. E {\bf 78}, 056707 (2008).
\bibitem{Batrouni2} G.G.~Batrouni and M.B.~Halpern, Phys. Rev. D {\bf 30}, 1775 (1984).
\bibitem{Feshbach1} V.G.~Rousseau and P.J.H.~Denteneer, Phys. Rev. A {\bf 77}, 013609 (2008).
\bibitem{Feshbach2} V.G.~Rousseau and P.J.H.~Denteneer, Phys. Rev. Lett. {\bf 102}, 015301 (2009).
\bibitem{Eckholt} Mar\'ia Eckholt and Tommaso Roscilde, Phys. Rev. Lett. {\bf 105}, 199603 (2010).
\bibitem{Andreev} A. F. Andreev and E. P. Bashkin, JETP 42, 164 (1975).
\end{thebibliography}
\end{document}